\documentclass[longauth]{aa} 

\usepackage{graphicx}
\usepackage{txfonts}
\usepackage{natbib}
\usepackage{booktabs}
\usepackage{xcolor} 
\usepackage{xspace}
\newcommand{\Msun}{\,M$_\odot$\xspace}     
\newcommand{\ergs}{\,erg\,s$^{-1}$\xspace} 

\newcommand{\apm}{APM\,08279+5255\xspace}
\newcommand{\grs}{GRS\,1734$-$292\xspace}
\newcommand{\icf}{IC\,4329A\xspace}

\newcommand{\mcgs}{MCG$-$06$-$30$-$15\xspace}
\newcommand{\mcge}{MCG+08$-$11$-$11\xspace}
\newcommand{\ngcn}{NGC\,985\xspace}

\begin{document}

   \title{Millimeter emission from supermassive black hole coronae}


   \author{S. del Palacio\inst{1}
          \and C.~Yang\inst{1}
          \and S.~Aalto\inst{1}
          \and C.~Ricci\inst{3,4}
          \and B.~Lankhaar\inst{1}
          \and S.~K\"onig\inst{2}
          \and J.~Becker Tjus\inst{1,5,6}
          \and M.~Magno\inst{14,15}
          \and K.\,L.~Smith\inst{14}
          \and J.~Yang\inst{2}
          \and L.~Barcos-Mu\~noz\inst{17,18}
          \and F.~Combes\inst{10}
          \and S.~Linden\inst{11} 
          \and C.~Henkel\inst{21,22} 
          \and J.\,G.~Mangum\inst{7}
          \and S.~Mart\'{\i}n\inst{24,25}
          \and G.~Olander\inst{1}  
          \and G.~Privon\inst{17}
          \and C.~Wethers\inst{1}
          \and A.-K.~Baczko\inst{1}
          \and R.\,J.~Beswick\inst{12}
          \and I.~Garc\'{\i}a-Bernete\inst{19}
          \and S.~Garc\'{\i}a-Burillo\inst{9}  
          \and E.~Gonz\'alez-Alfonso\inst{23}
          \and M.~Gorski\inst{27}
          \and M.~Imanishi\inst{8} 
          \and T.~Izumi\inst{20}
          \and S.~Muller\inst{2}
          \and Y.~Nishimura\inst{16}
          \and M.~Pereira-Santaella\inst{26}
          \and P.\,P.~van der Werf\inst{13}
          }

    \institute{Department of Space, Earth and Environment, Chalmers University of Technology, SE-412 96 Gothenburg, Sweden.\\
              \email{santiago.delpalacio@chalmers.se}  
    \and Department of Space, Earth and Environment, Chalmers University of Technology, Onsala Space Observatory, 43992 Onsala, Sweden.    
    \and Instituto de Estudios Astrofísicos, Facultad de Ingeniería y Ciencias, Universidad Diego Portales, Santiago 8370191, Chile 
    \and Kavli Institute for Astronomy and Astrophysics, Peking University, Beijing 100871, China          
    \and Theoretical Physics IV: Plasma-Astroparticle Physics, Faculty for Physics \& Astronomy, Ruhr University Bochum, 44780 Bochum, Germany 
    \and Ruhr Astroparticle And Plasma Physics Center (RAPP Center), Ruhr-Universit\"at Bochum, 44780 Bochum, Germany  
    \and National Radio Astronomy Observatory, 520 Edgemont Road, Charlottesville, VA 22903, USA 
    \and National Astronomical Observatory of Japan, 2-21-1, Osawa, Mitaka, Tokyo 181-8588, Japan 
    \and Observatorio Astron\'omico Nacional (OAN-IGN)-Observatorio de Madrid, Alfonso XII, 3, 28014-Madrid, Spain  
    \and Observatoire de Paris, LUX, Coll`ege de France, CNRS, PSL University, Sorbonne University, 75014, Paris, France 
    \and Steward Observatory, University of Arizona, 933 N Cherry Avenue, Tucson, AZ 85721, USA 
    \and Jodrell Bank Centre for Astrophysics, The University of Manchester, M13 9PL, UK 
    \and Leiden Observatory, Leiden University, PO Box 9513, 2300 RA Leiden, The Netherlands 
    \and George P. and Cynthia Woods Mitchell Institute for Fundamental Physics and Astronomy, Texas A\&M University, College Station, TX, 77845, USA 
    \and CSIRO Space and Astronomy, ATNF, PO Box 1130, Bentley WA 6102, Australia 
    \and Institute of Astronomy, The University of Tokyo, 2-21-1, Osawa, Mitaka, Tokyo 181-0015, Japan 
    \and National Radio Astronomy Observatory, 520 Edgemont Road, Charlottesville, VA 22903, USA 
    \and Department of Astronomy, University of Virginia, 530 McCormick Road, Charlottesville, VA 22903, USA 
    \and Centro de Astrobiolog\'ia (CAB), CSIC-INTA, Camino Bajo del Castillo s/n, E-28692 Villanueva de la Ca\~nada, Madrid, Spain 
    \and National Astronomical Observatory of Japan, 2-21-1 Osawa, Mitaka, Tokyo 181-8588, Japan 
    \and Max-Planck-Institut f\"ur Radioastronomie, Auf dem H\"ugel 69, D-53121 Bonn, Germany 
    \and Xinjiang Astronomical Observatory, Chinese Academy of Sciences, 830011 Urumqi, PR China 
    \and Instituto de F\'isica Fundamental, CSIC, Calle Serrano 123, E-28006, Madrid, Spain 
    \and European Southern Observatory, Alonso de C\'ordova, 3107, Vitacura, Santiago 763-0355, Chile 
    \and Joint ALMA Observatory, Alonso de C\'ordova, 3107, Vitacura, Santiago 763-0355, Chile 
    \and Instituto de F\'isica Fundamental, CSIC, Calle Serrano 123, 28006 Madrid, Spain 
    \and Center for Interdisciplinary Exploration and Research in Astrophysics (CIERA) Northwestern University, Evanston, IL 60208, USA 
    }

   \date{}

 
  \abstract
   {Active Galactic Nuclei (AGN) host accreting supermassive black holes (SMBHs). The accretion process can lead to the formation of a hot, X-ray emitting corona close to the SMBH capable of accelerating relativistic electrons. Observations in the millimetre band can probe its synchrotron emission.}
   {We intend to provide a framework to derive physical information of SMBH coronae by modelling their spectral energy distribution (SED) from radio to far infrared frequencies. We also explore the possibilities of deriving additional information from millimetre observations, such as the SMBH mass, and studying high-redshift lensed sources.}
   {We introduce a corona emission model based on a one-zone spherical region with a hybrid thermal and non-thermal plasma. We investigate in detail how the corona SED depends on different parameters such as size, opacity, and magnetic field strength. Other galactic emission components from dust, ionised gas and diffuse relativistic electrons are also included in the SED fitting scheme. We apply our code consistently to a sample of radio-quiet AGN with strong indications of a coronal component in the millimetre.}
   {The detected millimetre emission from SMBH coronae is consistent with having a non-thermal relativistic particle population with an energy density that is $\approx$0.5--10\% of that in the thermal plasma. This requires magnetic energy densities close to equipartition with the thermal gas, 
   and corona sizes of 60--250 gravitational radii.  
   The model can also reproduce the observed correlation between millimetre emission and SMBH mass when accounting for uncertainties in the corona size.}
   {The millimetre band offers a unique window into the physics of SMBH coronae, enabling the study of highly dust-obscured sources and high-redshift lensed quasars. Gaining a deeper understanding of the relativistic particle population in SMBH coronae can provide key insights into their potential multiwavelength and neutrino emission.
   } 


   \maketitle
%

\section{Introduction}

An Active Galactic Nucleus (AGN) is produced when a supermassive black hole (SMBH) accretes matter from its surrounding medium. The material around an SMBH is pulled by the strong gravitational field and spirals inward, forming an accretion disc. Processes related to turbulent magnetic viscosity in the accretion disc cause the material to lose angular momentum and fall towards the SMBH, releasing huge amounts of gravitational energy. The material then heats up, and radiates across a broad range of wavelengths \citep[e.g.][]{Balbus1998}. The accretion process is very efficient in converting gravitational potential energy into other forms, such as radiation, magnetic fields, and particle acceleration. Moreover, close to the innermost regions of the accretion disc, magnetic reconnection events can heat the gas to extremely high temperatures ($\sim$10$^9$~K), giving rise to a compact, hot region known as the corona \citep[e.g.][]{Haardt1991, Kamraj2022}. 

Most ($\sim$90\%) AGN emit only weakly at radio wavelengths and are therefore classified as radio-quiet (RQ). Although these AGNs lack a powerful radio jet \citep{Urry1995}, radio emission at 22~GHz is commonly detected \citep{Smith2020, Magno2025}. Interestingly, millimetre (mm) emission is also frequently observed in nearby RQ AGN, with most of the mm continuum flux originating from an unresolved core. Several studies have argued that this radiation is synchrotron emission produced in the corona, which also produces the hard X-ray emission above 2~keV through inverse Compton scattering of thermal disc photons, ubiquitously observed in AGN \citep[e.g.][]{Laor2008, Inoue2014, Behar2018, Panessa2019}. Recent studies strongly support this interpretation: high-resolution ($\sim$0.1--1") ALMA observations of nearby AGN revealed a tight correlation between X-ray emission and mm continuum at $\sim$200\,GHz \citep{Kawamuro2022} and at $\sim$100\,GHz \citep{Ricci2023}, as well as a correlation between the SMBH mass and the mm flux density at $\sim$230~GHz \citep{Ruffa2024}. In addition, correlated variability of X-ray and mm emission on year-timescales was found in NGC~1566 \cite{Jana2025}. Together, these results support the idea that the corona plays a key role in the mm emission from RQ AGN.

The X-ray corona is expected to be very compact, $r_\mathrm{c} = R_\mathrm{c}/R_\mathrm{g} \sim 10$--50, where $R_\mathrm{g} = G M_\mathrm{BH} /c^2$ is the gravitational radius and $M_\mathrm{BH}$ the mass of the SMBH \citep[e.g.][]{Fabian2015, Inoue2018}.
Nonetheless, the size of the corona is a topic of debate, with spectrotiming analysis of X-ray emission in Galactic X-ray binaries (XRBs) favouring dynamical coronae that vary in size from tens to hundreds of $R_\mathrm{g}$ \citep[e.g.][]{Karpouzas2021}. Even the geometry of the corona (lamp post, toroidal, spherical, sandwich-like) is unknown \citep[e.g.][and references therein]{Bambi2021}. Models for dynamical, outflowing coronae with sizes $r_c \sim 10$--$1000$ have also been suggested in the context of XRBs \citep{Kylafis2023}; this is remarkably consistent with the results of mm variability in the RQ AGN \icf \citep{Shablovinskaya2024}. On the topic of variability, we also highlight the results by \cite{Petrucci2023} showing rapid variability in mm and X-ray emission from MCG+08$-$11$-$11 within a day, and by \cite{Michiyama2024} revealing mm variability in a few days in \grs. The fast variability strongly supports that the mm emission comes from a compact region of $r_\mathrm{c} \sim 20$--$400$. Altogether, this shows that the size of the corona remains a topic of debate. 

In addition, the presence of ultra-relativistic particles in coronae is supported by both theoretical and observational evidence. Magnetic reconnection events, as well as other processes related to internal shocks and/or turbulence, can be responsible for accelerating relativistic particles out of thermal equilibrium \citep[\textit{non-thermal} particles; e.g.][]{Inoue2008, Beloborodov2017, Sironi2020, Groselj2024, Nattila2024}. In the case of a corona, the presence of extremely hot gas ($kT > 100$~keV) leads to the co-existence of both thermal and non-thermal relativistic particles \citep[e.g.][]{Ozel2000}, which we refer to as a `hybrid' plasma. Another independent proof of a non-thermal electron population in the corona is the hard X-ray/soft gamma-ray tail detected in the XRBs Cyg~X-1 \citep{Zdziarski2021} and MAXI~J1820+070 \citep{Cangemi2021}. A scenario involving (hadronic) cosmic rays in the corona is also compatible with the MeV polarised emission detected in Cyg~X-1 \citep{Romero2014}, and possibly with its sub-GeV and TeV emission \citep{Fang2024}. In addition, $\gamma$-ray emission in the  1--300~GeV energy range was detected in the stacked signal of 37~RQ AGN, which was interpreted to be produced in extended coronae \citep{Fermi2025}. Furthermore, several studies have suggested that significant neutrino emission can be produced by cosmic rays in SMBH coronae \citep[e.g.][]{Inoue2019, Murase2020, Kun2024, Fang2024}, consistent with the neutrino emission from NGC~1068 \citep{Inoue2020, Eichmann2022}. This scenario is particularly attractive because it can account for gamma-ray luminosity orders of magnitude lower than the neutrino luminosity, as the corona environment produces strong gamma-gamma absorption. 

A relatively new window to study SMBH corona is in the mm atmospheric bands. The relativistic electrons in the corona interact with the local magnetic fields and emit synchrotron radiation with a spectral energy distribution (SED) that peaks at around 100--300~GHz, depending on the plasma conditions in the corona and its non-thermal particle content \citep{Inoue2014, Inoue2019}. Although corona properties such as temperature and optical depth can be derived from X-ray observations \citep[e.g.][]{Fabian2015, Ricci2018}, the magnetic field strength and non-thermal particle population content can only be inferred by modelling the mm emission \citep[e.g.][]{Inoue2018}. 

Here, we present a framework for modelling the SEDs of RQ AGN. We pay special attention to the synchrotron emission from AGN coronae and its dependence on different physical parameters. We then apply this model consistently to a sample of objects for which good spectral coverage is available in order to derive the most relevant parameters of the coronae.

\section{Emission model} \label{sec:model}

The nuclear region of an RQ galaxy (inner $\sim$100\,pc) can contain large reservoirs of ionised gas, dust, cosmic rays, and an AGN. Thus, the observed continuum radio-to-submm SED of an RQ galaxy is the combination of several contributions, each with a distinct shape. We assume that these components are independent of each other and that the total emission is simply the sum of all components. The most relevant SED components that we consider are\footnote{All frequencies are given in the rest frame. For spectral indices we adopt the convention $S_\nu \propto \nu^{\alpha}$.}: 

\begin{enumerate}

    \item \textbf{Ionised gas:} The warm, ionised gas in the galaxy emits free--free (f--f) radiation that can be relevant at frequencies between $\sim 10$--100~GHz, where it presents a spectral index $\alpha=-0.1$ \citep[e.g.][]{Murphy2018}. The intensity of this emission is expected to scale with the amount of ionised gas and thus the star formation rate \citep[SFR;][]{Murphy2011}. In the simplest case, this emission can be parameterised as:
    \begin{equation}
    \centering
        S_\mathrm{ff}(\nu) = A_\mathrm{ff} \, \left( \dfrac{\nu}{\nu_0} \right)^{-0.1} \,, 
        \label{eq:ff}
    \end{equation} 
    where $A_\mathrm{ff}$ is a normalisation constant and $\nu_0$ is a reference frequency (fixed at 100~GHz). We note that, at low frequencies, this emission can actually drop as the medium becomes optically thick ($\alpha=2$ instead of $-$0.1). The turnover frequency depends most strongly on the density of the medium but also on its structure \cite[e.g.][]{Ramirez-Olivencia2022} 
    
    \item \textbf{A diffuse cosmic-ray electron population:} Its origin is likely related to the SFR, where supernova remnants and winds from massive stars accelerate cosmic rays that diffuse in the galaxy \citep[e.g.][]{Lacki2010, Kornecki2022}, or to extended jets \citep{Urry1995}. This component produces optically-thin synchrotron emission with a steep spectral index between $-0.5$ and $-1$ \citep[e.g.][]{Heesen2022, An2024}, dominating at low frequencies in the radio-cm range ($\lesssim 10$~GHz). In the simplest case, this emission can be parameterised as:
    \begin{equation}
    \centering
        S_\mathrm{sy}(\nu) = A_\mathrm{sy} \, \left( \dfrac{\nu}{\nu_0} \right)^{\alpha_\mathrm{sy}} \,, 
        \label{eq:sy}
    \end{equation}
    where $A_\mathrm{sy}$ is a normalisation constant, $\nu_0$ is a reference frequency (fixed at 100~GHz), and $\alpha_\mathrm{sy}$ is the intrinsic spectral index of the synchrotron emission. At low frequencies, the behaviour of this component can deviate from a simple power-law, most likely due to f--f absorption in the ionised gas that flattens the spectrum at lower frequencies \citep[e.g.][]{Kornecki2022, Dey2024}, consistent with what is observed in studies at $\nu < 1$~GHz of a large sample of galaxies \citep{Heesen2022, An2024}. If the ionised medium is inhomogeneous and clumpy, the absorption depends on the opacity and distribution of the clumps. This can result in significant absorption of the synchrotron emission, but also part of it can propagate between the clumps unaffected \citep[e.g.][]{Lacki2013, Ramirez-Olivencia2022}. 
    We refrain from including such effects in our model as it would introduce additional free parameters that do not assist us in our main goal, which is to isolate and study the corona component that peaks at much higher frequencies. 

    \item \textbf{Corona:} 
    The intrinsic synchrotron SED is a power-law with a negative spectral index ($\alpha \sim -0.5$ to $-1$), but lower frequency radiation is absorbed by the relativistic electrons via synchrotron-self absorption (SSA), producing a turnover in the SED that shifts to a positive spectral index $\alpha = 5/2$ \citep[e.g.][]{Margalit2021}. The position of the turnover frequency, $\nu_\mathrm{SSA} \approx 100$--300~GHz, depends on the plasma conditions in the corona, such as its density, size, and magnetic field strength (see forthcoming Sect.~\ref{sec:parameter_exploration}). This results in an SED with a rather pronounced peak at $\nu \sim \nu_\mathrm{SSA}$. As the main goal of this work is to derive the physical properties of the corona, in this case, we do not adopt a simple phenomenological model but rather compute the SED in a physically motivated way, detailed in Sect.~\ref{sec:corona_model}. We highlight that this region is extremely compact and can exhibit very high brightness temperatures (potentially up to $\sim10^9$~K if observed with a 10~$\mu$as resolution at 100~GHz, adopting as reference a flux density of $\sim$1~mJy).
    
    \item \textbf{Dust:} Its continuum emission dominates the SED at high frequencies ($\gtrsim 300$~GHz, depending on the source and the angular resolution of the observations). Within the Rayleigh-Jeans regime, this can be modelled as a modified-black body spectrum characterised by the frequency $\nu_{\tau_{1}}$ at which the dust opacity becomes equal to unity, and the index $\beta$ of the opacity coefficient $\kappa_\nu \propto \nu^\beta$ (with $\beta \approx 1$--2). At $\nu < \nu_{\tau_{1}}$ the SED is optically thin and has a spectral index $\alpha = 2+\beta$, while at $\nu > \nu_{\tau_{1}}$ it is optically thick and $\alpha = 2$. This is parameterised as:
    \begin{equation}
        S_\mathrm{d} = A_\mathrm{d} \, \left( \dfrac{\nu}{\nu_0} \right)^2 \, \left( 1 - e^{-\tau_\mathrm{d}} \right) \,, 
        \label{eq:dust}
    \end{equation}
    where $A_\mathrm{d}$ is a normalisation constant, $\nu_0$ is a reference frequency (fixed at 100~GHz), and $\tau_\mathrm{d} = (\nu/\nu_{\tau_{1}})^\beta$. 
    We note that the dust mass is $M_\mathrm{d} \propto \nu_{\tau_{1}}$ \citep{Draine2011}, so that an accurate determination of $\nu_{\tau_{1}}$ can help to estimate dust masses, although this would typically require a good spectral coverage between 400--1\,000~GHz that is challenging to obtain. We also note that, at the frequencies at which the dust becomes optically thick, it could also absorb the emission from the corona. However, this would depend on geometrical assumptions regarding the dust distribution along the line of sight, and it would only affect the optically thin part of the corona SED, which is difficult to observe as it is expected to be faint and typically below the dust continuum emission. We thus neglect this effect in our model fitting. 
    In addition, rapidly spinning dust grains can produce Anomalous Microwave Emission (AME) that peaks around $\nu \sim 30$~GHz, although in certain conditions, it can reach $\sim$100~GHz \citep{Dickinson2018, Murphy2020}. However, this component is faint and, in extragalactic contexts, it has only been detected in off-nuclear star-forming regions in a handful of nearby galaxies \citep[e.g.][]{Poojon2024, Fernandez-Torreiro2024}. It is thus unlikely that this component is bright in the sources of our sample, which in fact do not present a bump in their SED at (rest-frame) frequencies of $\sim$30~GHz (Sect.~\ref{sec:results}). For this reason, we do not include an AME emission component in our model.
\end{enumerate}

\subsection{Corona emission} \label{sec:corona_model}

We developed a corona emission model based on the one-zone model presented by \cite{Inoue2019}, in which the corona is a hybrid plasma with a thermal and a non-thermal particle population. We therefore use the numerical code from \cite{Margalit2021}\footnote{\url{https://github.com/bmargalit/thermal-synchrotron}, with the correction later introduced in \cite{Margalit2024}.} that computes the synchrotron SED of a hybrid plasma. We model the corona as a spherical region of size $R_\mathrm{c} = r_\mathrm{c} R_\mathrm{g}$, with $R_\mathrm{g}$ the gravitational radius and $r_\mathrm{c}$ an adimensional radius. The thermal plasma is characterised by its temperature $T_\mathrm{c}$ and number density $n_\mathrm{th,0}$. The latter is parameterised through the Thomson opacity $\tau_\mathrm{T}$ as $n_\mathrm{th,0} = \tau_\mathrm{T}/(\sigma_\mathrm{T}  R_\mathrm{c})$. The energy density in thermal electrons is $U_\mathrm{th,e} = a(\Theta) \, \Theta \, n_\mathrm{th,0} \, m_\mathrm{e} c^2$, where is $\Theta = k T_\mathrm{c} / (m_\mathrm{e} c^2)$ is an adimensional temperature and $a(\Theta) = (6 + 15\Theta)/(4 + 5\Theta)$ \citep{Margalit2021}.

Following \cite{Margalit2021}, for the non-thermal population we define a minimum Lorentz factor $\gamma_\mathrm{min}(\Theta) = 1+a(\Theta)\,\Theta$, such that at energies above $E_\mathrm{min} = \gamma_\mathrm{min} m_\mathrm{e} c^2$ the particle energy distribution is:
\begin{equation}
N(E) \propto \begin{cases}
    E^{-p+1}, \quad t_\mathrm{sy} \leq t_\mathrm{dyn} \\
    E^{-p}, \quad t_\mathrm{sy} > t_\mathrm{dyn} \end{cases} .   
\end{equation}
Here, $p$ is the spectral index of the electron distribution without cooling, $t_\mathrm{dyn}$ is the characteristic dynamical timescale for which we adopt the free-fall time $t_\mathrm{dyn}=R_\mathrm{c}/v_\mathrm{ff}$ \citep[with $v_\mathrm{ff} = \sqrt{2GM/R_\mathrm{c}}$ the free-fall velocity;][]{Inoue2019}, and $t_\mathrm{sy}$ is the synchrotron cooling timescale. We highlight that the values of $t_\mathrm{sy}$ and $t_\mathrm{dyn}$ are calculated self-consistently in the model for any given set of parameters ($M_\mathrm{BH}$, $r_\mathrm{c}$, etc.).

The next step is to define the non-thermal electron population in the corona and the magnetic field. For this, additional free parameters have to be introduced. One sensible way to do this is to use the energy density in thermal electrons as a reference. We define the ratio between the energy densities in the magnetic field and thermal electrons $\epsilon_B = U_B/U_\mathrm{th,e}$, and the ratio between the energy density in non-thermal electrons and thermal electrons $\delta = U_\mathrm{nt,e}/U_\mathrm{th,e}$. To constrain the model further, we assume $\delta < 1$, as otherwise the non-thermal electrons would dominate in the corona, which does not seem to be the case \citep[e.g.][]{Zdziarski2021}; we further note that \cite{Inoue2018} fixed this parameter to 0.04. Another consideration is that magnetic fields are believed to play a major role in heating the corona and accelerating relativistic particles \citep[e.g.][]{Beloborodov2017, Sironi2020}, and therefore it should be the case that $\epsilon_B$ is of order unity and $\epsilon_B > \delta$. We can thus impose an energy condition $U_B = \eta U_\mathrm{nt}$, where $U_\mathrm{nt}$ is the total energy in non-thermal particles. It is usual to write this as $U_\mathrm{nt} = U_\mathrm{nt,e} + U_\mathrm{nt,p} = (1 + \xi_\mathrm{e,p}) \, U_\mathrm{nt,e}$, where $\xi_\mathrm{e,p}$ is the ratio between the energy density in relativistic protons and electrons. With these definitions, we can rewrite $\epsilon_B = \eta \, \delta \, ( 1+\xi_\mathrm{e,p})$. Either because relativistic protons are easier to accelerate or because relativistic electrons cool faster than protons, the energy density in non-thermal protons is larger than that of electrons \citep[e.g.][]{Eichmann2022}, and therefore we fix $\xi_\mathrm{e,p}=40$ \citep[roughly corresponding to $\sqrt{m_\mathrm{p}/m_\mathrm{e}}$; e.g.][]{Merten2017}. We additionally fix $\eta=1$, corresponding to an energy equipartition condition between non-thermal particles and magnetic fields. In this way, we remove a free parameter by the condition $\epsilon_B = 41 \, \delta$.

A related parameter commonly used to characterise the plasma conditions is the $\sigma$ parameter defined as $\sigma = B^2 / (4\pi n_\mathrm{th,e} m_\mathrm{e} c^2)$. Neglecting some factors of order unity, one can approximate $\sigma \approx \epsilon_B \, \Theta$, which helps to understand the dependence of $\sigma$ on the model parameters. 
In particular, for a highly magnetised plasma, one expects $\sigma$ to be of the order of unity. Similarly, this can be compared with the plasma $\beta_B$ parameter defined as the ratio of thermal and magnetic pressures, $\beta_B = P_\mathrm{th}/P_B$, which can be rewritten as\footnote{If a two-temperature plasma is assumed for the corona, for $r_\mathrm{c} \sim 140$ we expect an ion temperature of $T_\mathrm{i} \sim 1.1$\,MeV \citep[e.g. Eq.~1 in][]{Inoue2024}, and thus a total thermal pressure of $P_\mathrm{th} \sim (1 + T_\mathrm{i}/T_\mathrm{c}) \, U_\mathrm{th,e} / a(\Theta) \sim 5 U_\mathrm{th,e}$.} $\beta_B \sim 5/\epsilon_B$. 

Once the thermal and non-thermal particle populations have been defined, together with the physical properties of the corona, the code calculates the synchrotron SED, taking into account the emissivity and opacity (SSA) from both populations. In this way, the synchrotron SED (including $\nu_\mathrm{SSA}$) is self-consistently calculated for any set of parameters. 

We finally convert the specific luminosities to flux densities in order to compare them with observed SEDs. For this, we use the \texttt{Planck18} library available in \texttt{astropy} to convert $z$ to a luminosity distance, and the rest-frame frequencies to the observed frame as $\nu_\mathrm{obs}=\nu/(1+z)$. Finally, for studying lensed sources (as explored in Sect.~\ref{sec:high_z}), we introduce a multiplicative amplification factor $\mu$ to the intrinsic luminosity. 

\begin{table*}
\caption{Summary of all the parameters in the model described in Sect.~\ref{sec:model}. $^\dagger$Fixed to the literature value when possible.}       
\label{table:parameters}                        
\centering                        
\begin{tabular}{l c c c}          
\hline\hline                       
Parameter (units)                               & Symbol                & Values             & Comments \\ 
\hline                          
Redshift of the source                          & $z$                   & $>0$                  &  Fixed to the literature value \\    
Normalisation of the f--f component             & $A_\mathrm{ff}$       & [$10^{-3}$, $10^{1}$] &  Defined in Eq.~\ref{eq:ff} \\
Normalisation of the diffuse synchrotron component & $A_\mathrm{sy}$    & [$10^{-3}$, $10^{1}$] &  Defined in Eq.~\ref{eq:sy} \\ 
Spectral index of the diffuse synchrotron component & $\alpha_\mathrm{sy}$ & [$-2$, $-0.45$]     &  Defined in Eq.~\ref{eq:sy} \\ 
Normalisation of the dust component             & $A_\mathrm{d}$        & [$10^{-3}$, $10^{1}$] &  Defined in Eq.~\ref{eq:dust} \\
Dust opacity index                              & $\beta$               & [1, 2]                &  Defined in Eq.~\ref{eq:dust}. Default value: 1.78 \\
Frequency at which $\tau_\mathrm{d} = 1$ (GHz)  & $\nu_{\tau_1}$        & [300, 1300]           &  Defined in Eq.~\ref{eq:dust}. Default value: 700  \\
\hline \addlinespace[1pt] 
Mass of the supermassive black hole             & $M_\mathrm{BH}$       & [$10^{6}$, $10^{9}$]  & $\dagger$  \\ 
Corona radius ($r_g$)                           & $r_\mathrm{c}$        & [10, 500]             &   \\ 
Corona temperature (keV)                        & $k\,T_\mathrm{c}$     & [10, 200]             & Default value: 166$^\dagger$  \\
Corona Thomson opacity                          & $\tau_\mathrm{T}$     & [0.01, 10.0]          & Default value: 0.25$^\dagger$  \\
Ratio of non-thermal particle energy density in the corona & $\delta$            & [$10^{-4}$, $10^{0}$] &  \\ 
Spectral index of the injected non-thermal particles & $p$              & (2, 4]                & Default value: 2.7 \\ 
Magnetic field energy density in the corona     & $\epsilon_B$          & [$10^{-3}$, $10^{1}$] & Default value: $41\,\delta$ \\ 
\hline              
\end{tabular}
\end{table*}

\subsection{Parameter fitting} \label{sec:parameter_fitting}

\begin{figure*}
    \centering
    \includegraphics[width=\linewidth]{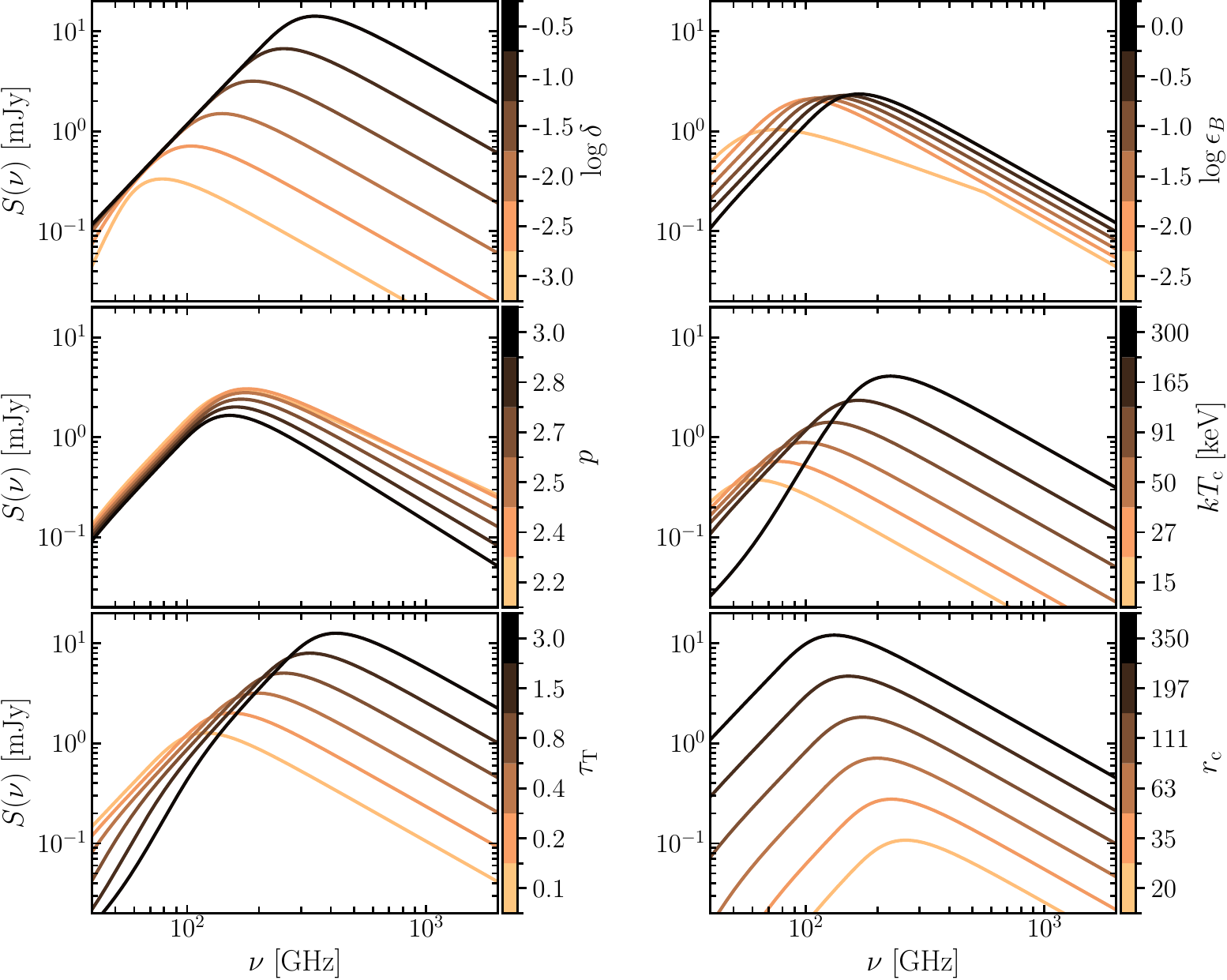} 
    \caption{Corona SEDs varying only one parameter at a time, as indicated in the colourbar of each panel. The fiducial values are $D = 100$\,Mpc, $M_\mathrm{BH} = 10^8$\Msun, $\tau_\mathrm{T}=0.25$, $kT_\mathrm{c}=166$\,keV, $r_\mathrm{c}=130$, $\log{\epsilon_B}=0$, $\log{\delta}=-1.5$, and $p=2.7$. More details are discussed in Sect.~\ref{sec:parameter_exploration}.}
    \label{fig:corona_seds}
\end{figure*}

We developed a code\footnote{\url{https://github.com/santimda/sed_fitter_BHcorona}} that calculates the total SED from the physical coronal model, together with the phenomenological components described in Sect.~\ref{sec:model}. We use the Python package \texttt{Bilby} \citep{bilby} for fitting the SED using the standard sampler for Markov Chain Monte Carlo \texttt{emcee} \citep{emcee}. One key issue is dealing with observations with different angular resolutions. The corona is extremely compact, so its emission should be point-like in all observations. The other galactic components are most likely extended and inhomogeneous, so it is not possible to scale the observed flux density with the beam size in a reliable way. For this reason, we use beam-matched data whenever possible, even if that requires compromising angular resolution. The main drawback is that the corona emission is not very bright, and so it can be well below the continuum of other galactic components in observations with poor angular resolution that collect too much diffuse emission, making it undetectable. For this reason, we filter out observations with coarse angular resolution (typically $> 1"$, corresponding to linear scales $>$100~pc for galaxies at distances beyond 20~Mpc) and consider them only as strict upper limits (ULs) for the total flux density. Similarly, when it comes to observations with significantly higher angular resolution than the rest of the dataset (e.g., from very large baseline interferometry---VLBI), we consider them only as strict lower limits (LLs) of the total flux density. In both cases, we leave a 1-$\sigma$ margin to be more conservative about using points as strict ULs or LLs, that is, for a measured flux of $S_\nu \pm \Delta S_\nu$, we choose $S_\mathrm{UL} = S_\nu + \Delta S_\nu$ and $S_\mathrm{LL} = S_\nu - \Delta S_\nu$. 

To reduce the number of free parameters in the model, we can use the parameterisation for $\epsilon_B(\delta)$ described in Sect.~\ref{sec:corona_model}. In addition, in cases where there is no direct determination of $\tau_\mathrm{T}$ and $T_\mathrm{c}$ from X-ray data, we adopt the parameterisation for $T_\mathrm{c}(\tau_\mathrm{T})$ from \cite{Tortosa2018}. As a reference, this corresponds to $kT_\mathrm{c}\simeq 166$~keV for the most typical case of $\tau_\mathrm{T} \simeq 0.25$ \citep{Ricci2018}, while for $\tau_\mathrm{T} \simeq 1$ we get $kT_\mathrm{c}\simeq 63$~keV. We summarise all the model parameters and the range of values that they can take in Table~\ref{table:parameters}. We note that for quantities that can span several orders of magnitude, we fit the logarithm of the quantity.

This code has already been used to model the SEDs of \icf \citep{Shablovinskaya2024} and NGC~1068 \citep{Mutie2025}. As a sanity check, we verified that the sizes derived with this approach agree well with the simple analytical estimate from Eq.~4 in \cite{Shablovinskaya2024}. Additionally, we confirmed that our results are consistent with those reported by \cite{Inoue2018} when fixing the same parameters as in their work.

Throughout this work, the quoted errorbars correspond to the 16th--84th percentile range (the $1\sigma$ confidence intervals for Gaussian distributions) and are derived from the posterior distribution. We also assume a conservative flux uncertainty to account for systematic errors in absolute flux calibrations, such that the total errorbar is $\Delta S_\nu = \sqrt{\Delta S_\mathrm{sta}^2 + \Delta S_\mathrm{sys}^2}$, with $\Delta S_\mathrm{sta}$ the statistical uncertainty and $\Delta S_\mathrm{sys} = f_\mathrm{sys}\,S_\nu$ the assumed systematic error. The factor $f_\mathrm{sys}$ is fixed to 5\% for all observations, except for ALMA observations at B6--8 (212--500~GHz) and B9--10 (600--900~GHz), for which it is 10\% and 20\%, respectively\footnote{\url{https://almascience.nrao.edu/proposing/proposers-guide#autotoc-item-autotoc-82}}.

\section{Results} \label{sec:results}

\subsection{Parameter exploration} \label{sec:parameter_exploration}

We explore the corona SED response to the different model parameters in order to gain some intuition of the interpretation of the observed SEDs. For this we consider a reference case with the following parameters: $D = 100$\,Mpc, $M_\mathrm{BH} = 10^8$\Msun, $\tau_\mathrm{T}=0.25$, $kT_\mathrm{c}=166$\,keV, $r_\mathrm{c}=140$, $\epsilon_B=1.0$, $\delta=0.01$, and $p=2.7$. First, we vary one parameter at a time. The results are shown in Fig.~\ref{fig:corona_seds}. Some highlights are:
\begin{itemize}
    \item Higher $\delta$ values mean more relativistic electrons and, therefore, lead to a higher emission and a higher opacity. Below $\nu_\mathrm{SSA}$ the emission is optically thick, so that a larger number of non-thermal electrons does not increase the flux, which remains constant for all $\delta$ values. The value of $\nu_\mathrm{SSA}$ does however increase with increasing $\delta$. 
    \item The value of $\epsilon_B$ sets the magnetic field strength. Higher magnetic fields lead to more SSA, and therefore $\nu_\mathrm{SSA}$ depends strongly on this parameter. Moreover, for very small magnetic fields ($\epsilon_B < 0.01$), the electrons do not cool efficiently ($t_\mathrm{sy} > t_\mathrm{dyn}$), and the flux in the optically thin part of the SED varies significantly with $\epsilon_B$. For values of $\epsilon_B > 0.01$ the dependence of the SED with this parameter is, however, rather weak.
    \item The value of $p$ affects mostly the slope in the optically thin part of the spectrum. Thus, observational restrictions on this parameter can only be obtained with observations at frequencies significantly higher than $\nu_\mathrm{SSA}$. The frequency $\nu_\mathrm{SSA}$ itself does not vary significantly with $p$. 
    \item Increasing the corona temperature increases $U_\mathrm{th,e}$, and therefore also $U_\mathrm{nt,e}$ and $U_B$ for fixed values of $\delta$ and $\epsilon_B$. Thus, increasing $T_\mathrm{c}$ has a similar impact as increasing $\delta$ and $\epsilon_B$. The main difference occurs for $k T_\mathrm{c} \sim 300$~keV, at which point the thermal electrons play a significant role in further absorbing the emission below $\nu_\mathrm{SSA}$, although their emission is negligible \citep[we refer to][for a scenario in which the thermal electrons can have an even more dominant role]{Margalit2021}. For completeness, we explore in more detail the emission from thermal electrons in Appendix~\ref{app:thermal_synchrotron}.
    \item The Thomson opacity defines the density of thermal particles and, therefore, impacts $U_\mathrm{th,e}$. The increase in $\tau_\mathrm{T}$ therefore has a similar impact as the increase in $k T_\mathrm{c}$.
    \item For a fixed $M_\mathrm{BH}$, the value of $r_\mathrm{c}$ determines the size of the corona. A bigger corona produces more emission, and it is also more transparent as it is more diluted (as $n_\mathrm{th,0} \propto \tau_\mathrm{T} \, r_\mathrm{c}^{-1}$). Thus, $\nu_\mathrm{SSA}$ decreases with an increasing $r_\mathrm{c}$. This is the only parameter for which an increase in the total peak flux is accompanied by a decrease in $\nu_\mathrm{SSA}$, and it is, therefore, a key parameter in the SED fitting to get the correct peak position and flux. 
\end{itemize}

\begin{figure}
    \centering
    \includegraphics[width=\linewidth]{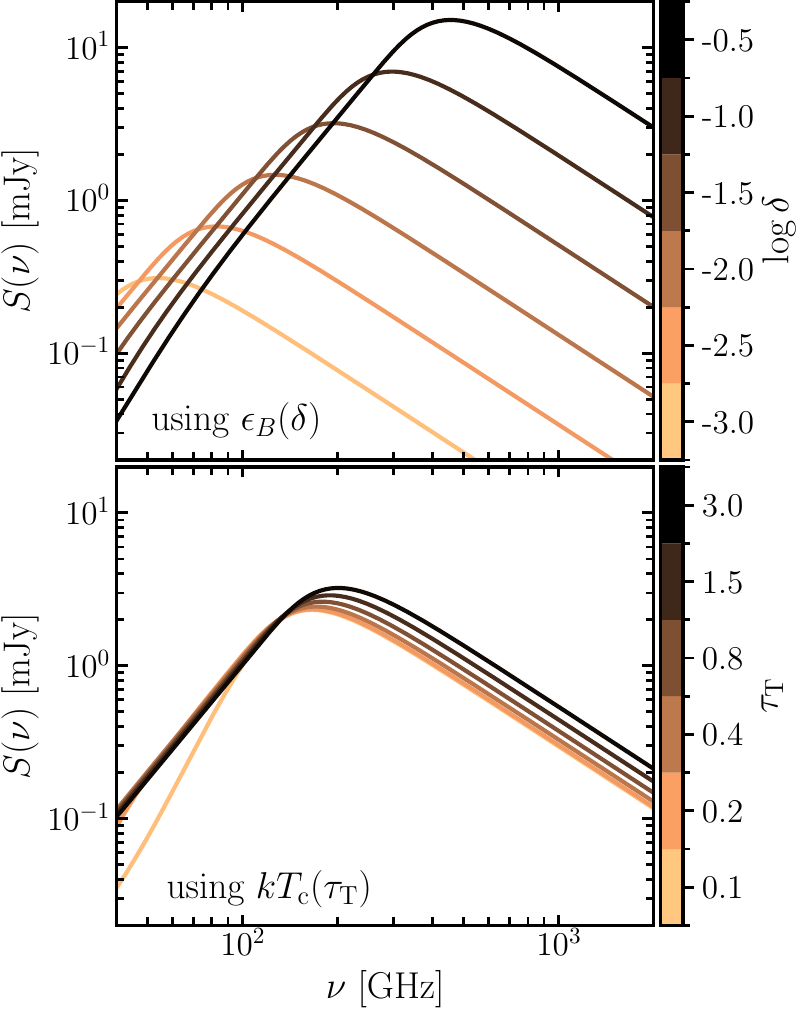}\\
    \caption{Same as Fig.~\ref{fig:corona_seds} but using the relations $\epsilon_B(\delta)$ (top panel) and $T_\mathrm{c}(\tau_\mathrm{T})$ (bottom panel) discussed in Sect.~\ref{sec:parameter_exploration}. }
    \label{fig:corona_seds_tied_parameters}
\end{figure}

As discussed in Sect.~\ref{sec:corona_model}, we can reduce the number of free parameters by assuming an energy condition of the form $\epsilon_B = \eta \, \delta \, ( 1+\xi_\mathrm{e,p}) = 41 \, \delta$ and the observational relation for $T_\mathrm{c}(\tau_\mathrm{T})$ from \cite{Tortosa2018}. The results are shown in Fig.~\ref{fig:corona_seds_tied_parameters}. We can summarise them as:
\begin{itemize}
    \item Increasing $\delta$ (with $\epsilon_B(\delta)$) leads to a significant increase in both $\nu_\mathrm{SSA}$ and the peak flux. 
    \item Changes in $\tau_\mathrm{T}$ (with $T_\mathrm{c}(\tau_\mathrm{T})$) have negligible effects in the SED. This is because the SED depends in similar ways on $\tau_\mathrm{T}$ and $T_\mathrm{c}$, and $\tau_\mathrm{T}$ and $T_\mathrm{c}$ are anti-correlated, thus compensating the changes in the SED. This means that, in general, we do not expect the pairs of values of $\tau_\mathrm{T}$ and $T_\mathrm{c}$ to have a significant impact on the SED. 
\end{itemize}

\begin{figure}
    \centering    
    \includegraphics[width=\linewidth]{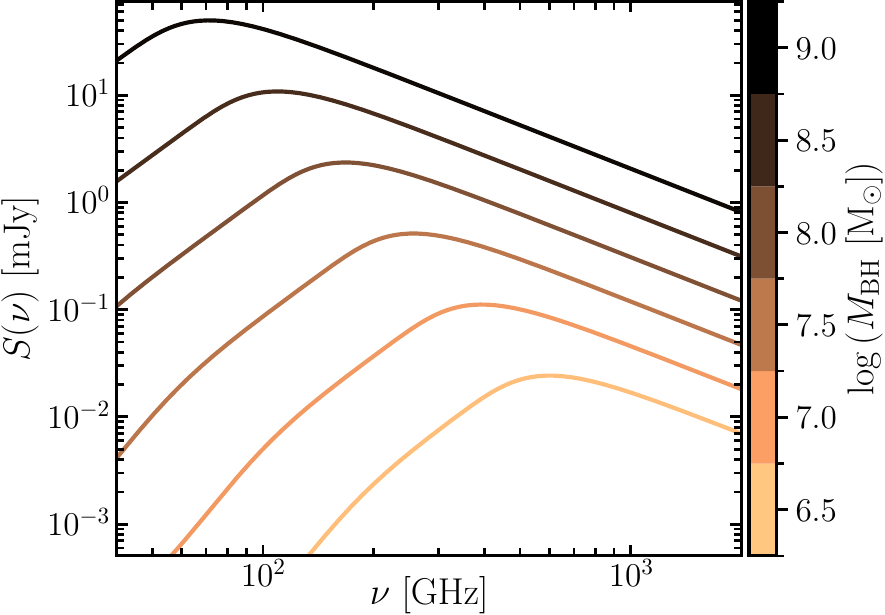}
    \caption{Corona SEDs for different values of the SMBH mass. As $M_\mathrm{BH}$ increases, the total flux increases and $\nu_\mathrm{SSA}$ decreases.}
    \label{fig:corona_sed_M}
\end{figure}

On a final note, the SMBH mass, $M_\mathrm{BH}$, has a very strong impact on the SED as it affects the size of the corona (for a fixed $r_\mathrm{c}$) and, to a lesser extent, the non-thermal electron distribution (via $v_\mathrm{ff}$). In Fig.~\ref{fig:corona_sed_M}, we show corona SEDs for a broad range of SMBH masses. The behaviour is similar to that seen for variations in $r_\mathrm{c}$, i.e., a decreasing $\nu_\mathrm{SSA}$ and an increasing flux with an increasing $M_\mathrm{BH}$. We show that for the most massive SMBHs, we expect the corona emission to peak at a frequency $\nu_\mathrm{p} < 100$~GHz, whereas for low-mass SMBHs, the peak should fall at significantly higher frequencies ($\nu_\mathrm{p} > 400$~GHz). More specifically, the dependence of the peak frequency with the black hole mass in Fig.~\ref{fig:corona_sed_M} is $\nu_\mathrm{p} \propto M_\mathrm{BH}^{-0.37}$.

\subsection{Studying sources at high redshift} \label{sec:high_z}

The detection of mm coronal emission has so far been almost exclusively limited to local AGN due to sensitivity limitations. This can potentially be overcome in the case of high-$z$ lensed quasars with high amplification factors $\mu$ and large $M_\mathrm{BH}$. Remarkably, a hint of such coronal emission was found in \apm, an AGN at $z\simeq3.9$ hosting an SMBH with $M_\mathrm{BH} \approx 10^{9.8}$\Msun \citep{Yang2023b}, and a clear detection was recently reporeted in the lensed quasar RXJ1131$-$1231, located at $z = 0.658$ \citep{Rybak2025}.

\begin{figure}
    \centering    
    \includegraphics[width=\linewidth]{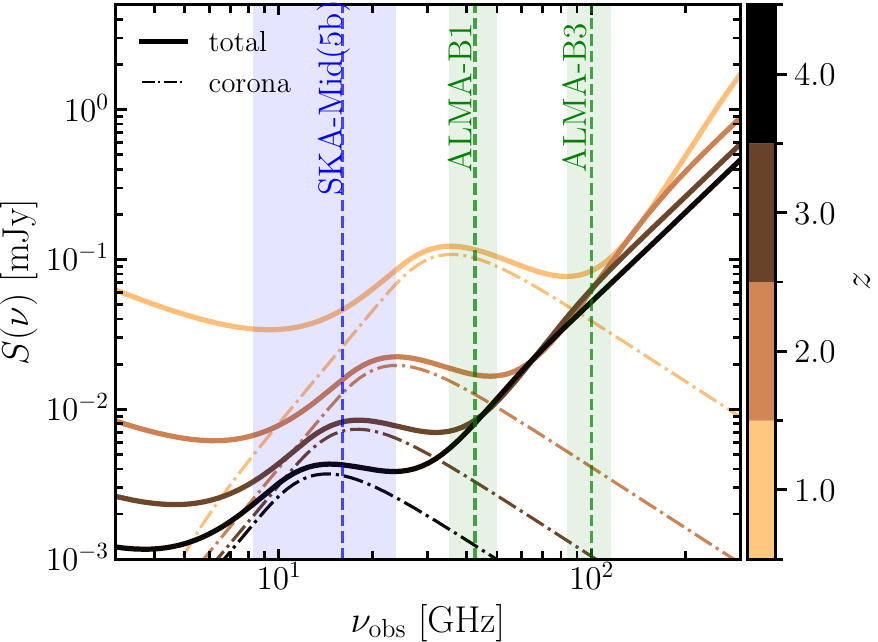}
    \caption{Example SEDs for a source with $M_\mathrm{BH}= 10^9$\Msun and $\mu=10$ at different redshifts $z$. Arbitrary synchrotron and dust components were added to mimic more realistic SEDs. The frequency ranges for highly sensitive observatories are also shown.}
    \label{fig:corona_sed_z}
\end{figure}

In Fig.~\ref{fig:corona_sed_z}, we show the corona SEDs for a source with $M_\mathrm{BH}= 10^9$\Msun and $\mu=10$ located at different $z$ values. The remaining parameters are the same as those used in Fig.~\ref{fig:corona_sed_M}. For sources at $z>1$, the peak of the SED falls below 40~GHz, and their maximum flux density falls below 0.1~mJy, thus making their observation unfeasible for ALMA. Instead, these are good candidates for the upcoming Square Kilometer Array (SKA\footnote{\url{https://www.skao.int/en}}). For conditions similar to those assumed here, we predict that the SKA-Mid could detect such sources up to $z \sim 4$ with a $\sim$1~h integration (reaching sensitivities of $\sim$1~$\mu$Jy). 

\subsection{Analysis of the sample sources}\label{sec:results_sources}

We apply our model and fitting scheme to a sample of seven RQ AGNs that includes all sources for which strong indications of coronal activity have been reported in the literature (\grs, \icf, \mcge, \ngcn and NGC~1068), together with two sources for which high-resolution data from radio-cm to (sub)mm is available (\mcgs and NGC~3227). For all sources, we do a detailed bibliographic search for published flux densities and then include additional data from unpublished archival observations from various surveys. Namely, we search for archival data with the Karl G. Jansky Very Large Array (VLA)\footnote{\url{https://www.vla.nrao.edu/cgi-bin/nvas-pos.pl}}, and surveys accessed via the CIRADA image web cutout service\footnote{\url{http://cutouts.cirada.ca/}}, in particular the VLASS \citep{Lacy2020}, NVSS \citep{NVSS}, and RACS \citep{Hale2021}. 
In Appendix~\ref{app:sample}, we provide a short description of each source, and in Tables~\ref{table:grs_fluxes}-\ref{table:ngc3227_fluxes} we detail the data used, including the flux densities and relevant angular sizes. 

\begin{figure*}
    \centering
    \includegraphics[width=0.485\linewidth]{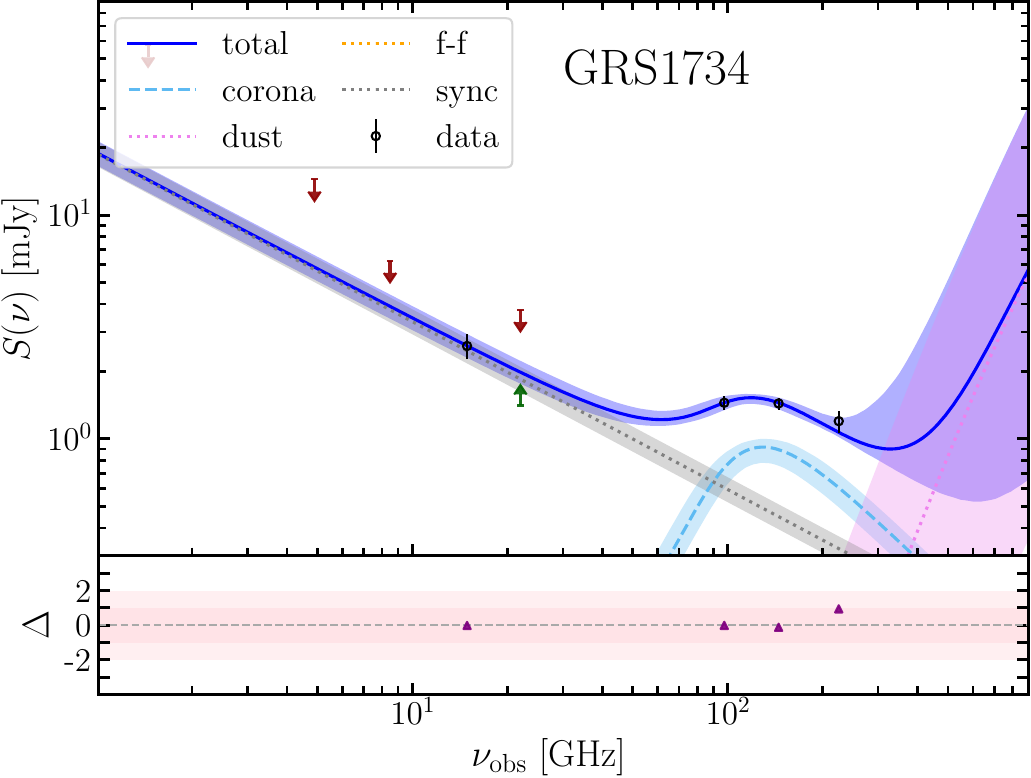} \hfill
    \includegraphics[width=0.485\linewidth]{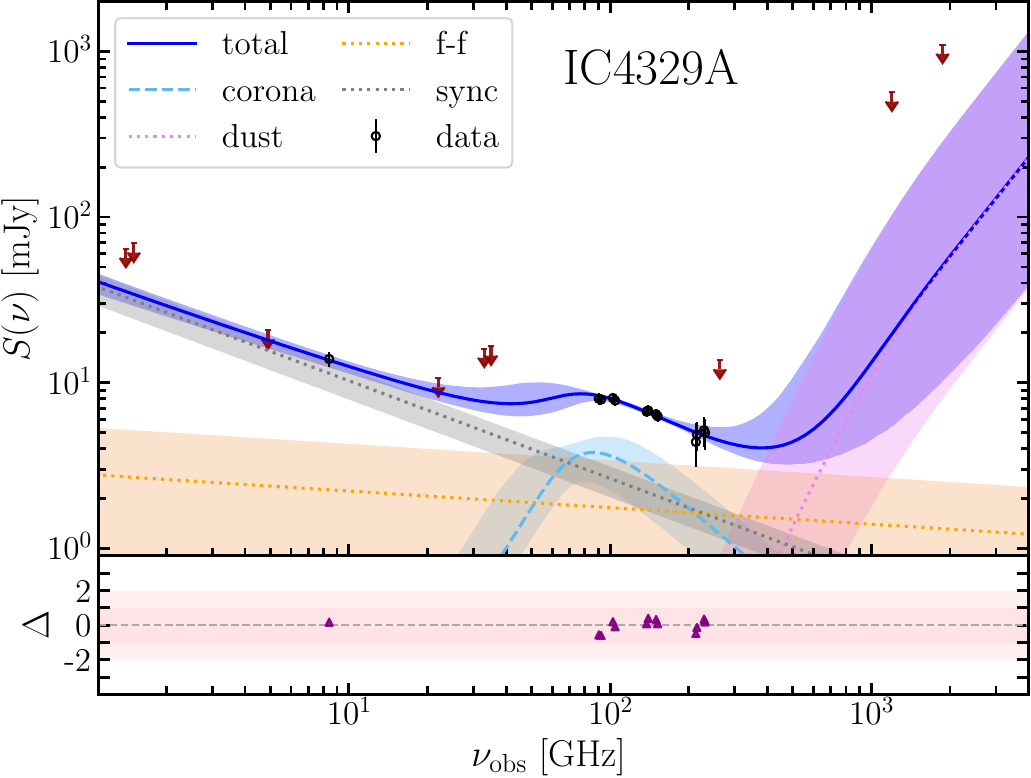}\\
    \includegraphics[width=0.485\linewidth]{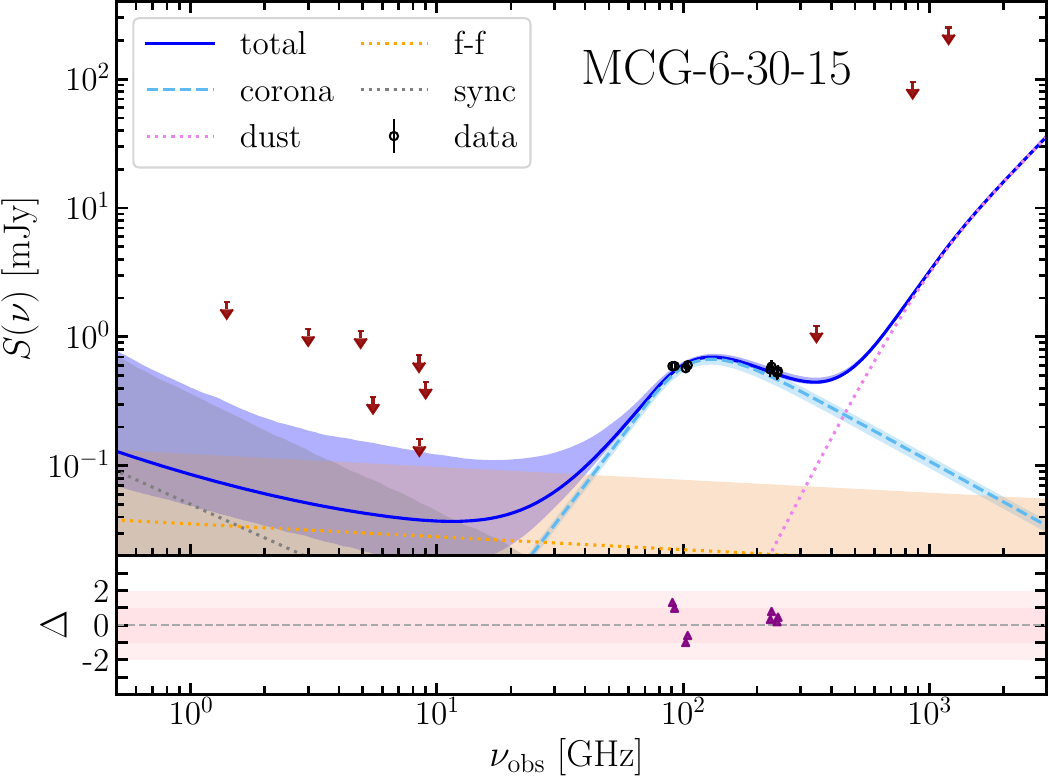} \hfill
    \includegraphics[width=0.485\linewidth]{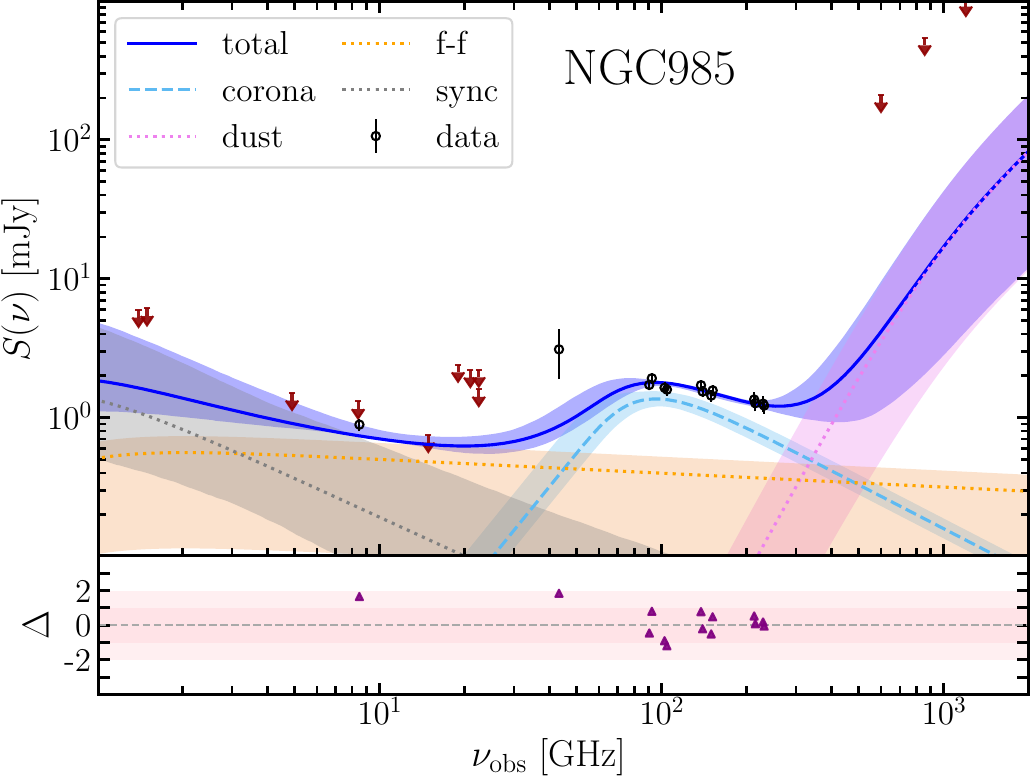} \\
    \includegraphics[width=0.485\linewidth]{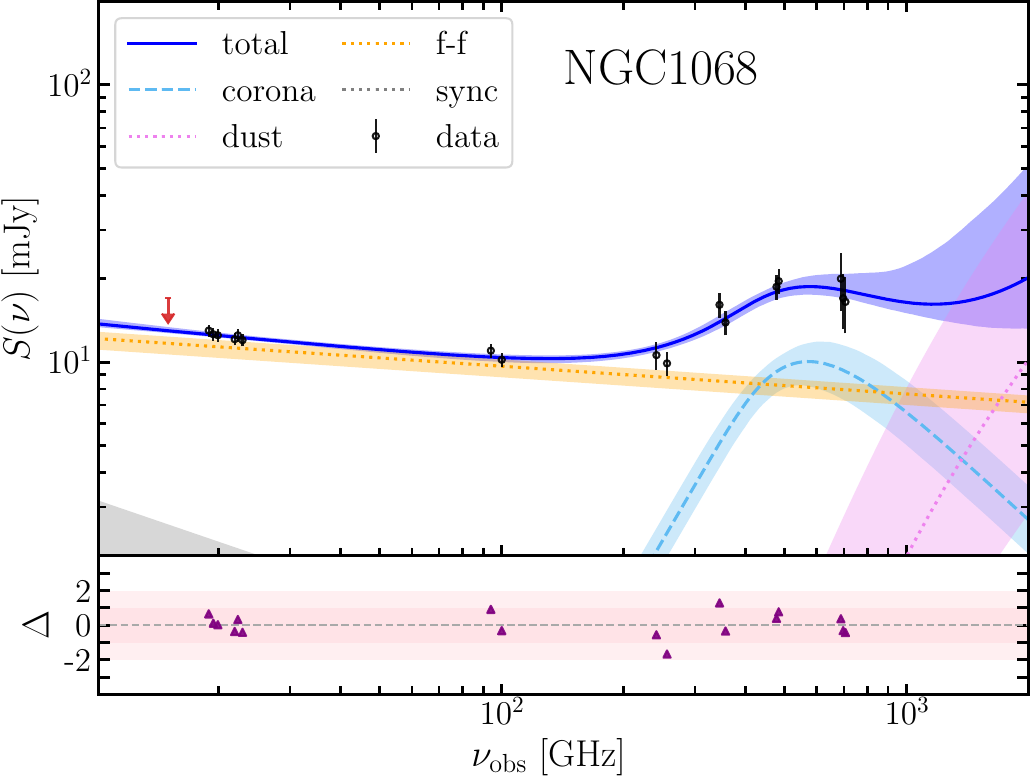} \hfill
    \includegraphics[width=0.485\linewidth]{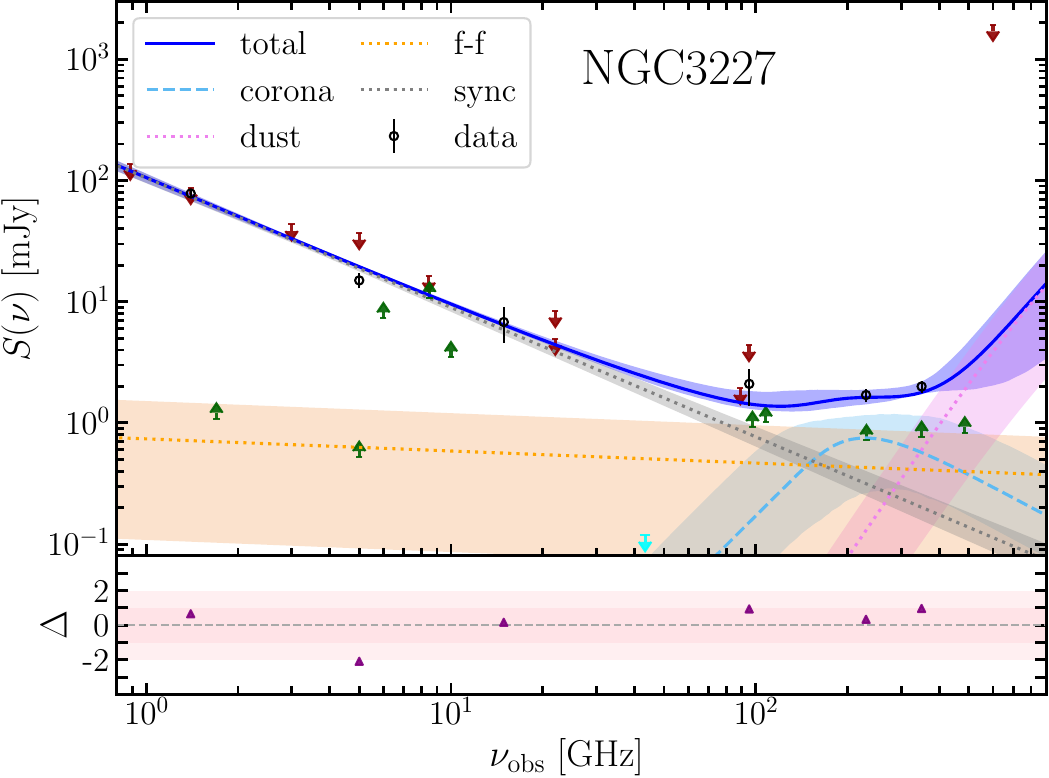} 
    \caption{Multifrequency SEDs and model fitting of the RQ galaxies in the sample. Data from low-resolution observations are treated as upper limits and shown with red arrows, whereas data from very-high-resolution observations are treated as lower limits and shown with green arrows (Sect.~\ref{sec:parameter_fitting}). Further details are provided in Appendix~\ref{app:sample}.
    For NGC~3227, we also show an upper limit at 43~GHz (cyan arrow) not used in the fitting due to its very high angular resolution, but the compact corona emission should still remain below this value for consistency.
    } 
    \label{fig:sources_seds_non-cons}
\end{figure*}

The SEDs of six sources and their fit are shown in Fig.~\ref{fig:sources_seds_non-cons}. We leave \mcge out of this part of the analysis, since its corona component is not constrained by the currently available data (Fig.~\ref{fig:SED_mgce}). More details about the SED fitting procedure in each source are discussed in Appendix~\ref{app:sample}, and the corresponding fit posteriors are provided in Figs.~\ref{fig:grs_cornerplot}--~\ref{fig:ngc3227_cornerplot}. 

\begin{figure}
    \centering
    \includegraphics[width=\linewidth]{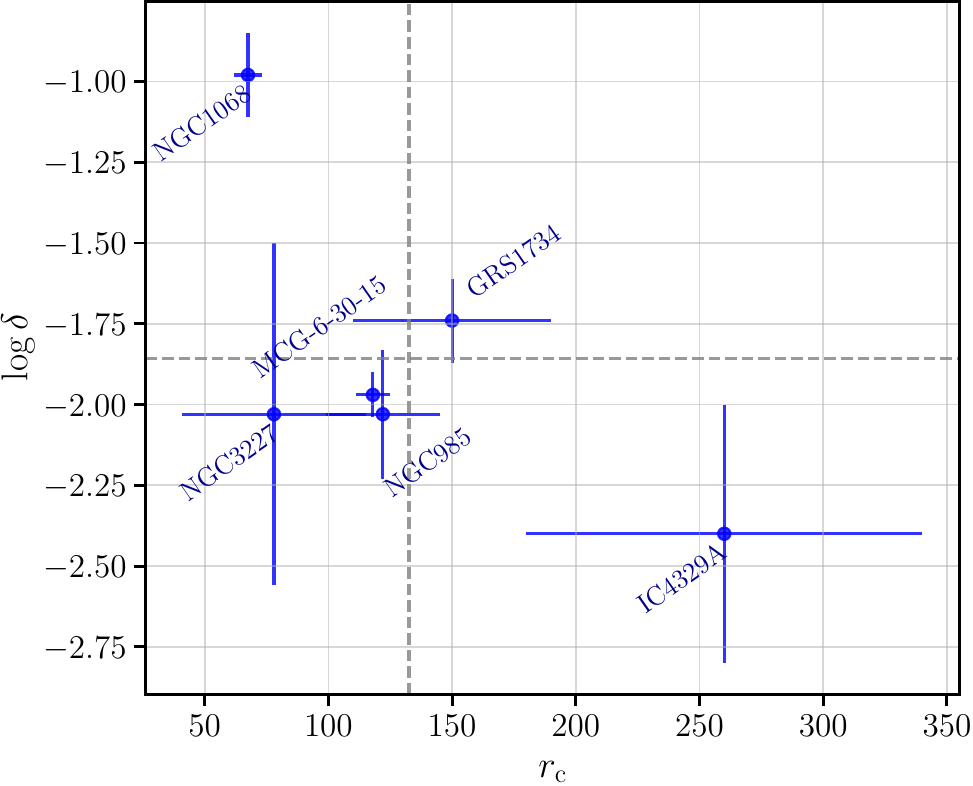} 
    \caption{Non-thermal particle content and corona sizes obtained for the sources in the sample via the SED fitting (with the parameters defined as in Sects.~\ref{sec:corona_model} and \ref{sec:parameter_fitting}). The grey dashed lines mark the median values $r_\mathrm{c}=133$ and $\log{\delta}=-1.86$. For the sources \grs and \icf, the broad range in $r_\mathrm{c}$ reflects the range in sizes derived in different epochs.}
    \label{fig:r_vs_delta}
\end{figure}

We summarize the results of the SED fitting of the two relevant corona parameters, $r_\mathrm{c}$ and $\delta$, in Fig.~\ref{fig:r_vs_delta}. The parameters range in the values $r_\mathrm{c} = 60$--250 and $\delta=10^{-3}$--10$^{-1}$, with median values of $r_\mathrm{c} \approx 133$ and $\log{\delta} \approx -1.86$ ($\delta \approx 1.4$\%). The derived values of $B$ range between 10--150~G, with a median of $\approx$20~G. 
We also find values of $\sigma$ ranging between 0.01--1, with a mean value of $\lesssim$0.1, roughly consistent with the assumption of a significantly magnetised medium ($\sigma \sim 1$). 

\begin{figure}
    \centering
    \includegraphics[width=\linewidth]{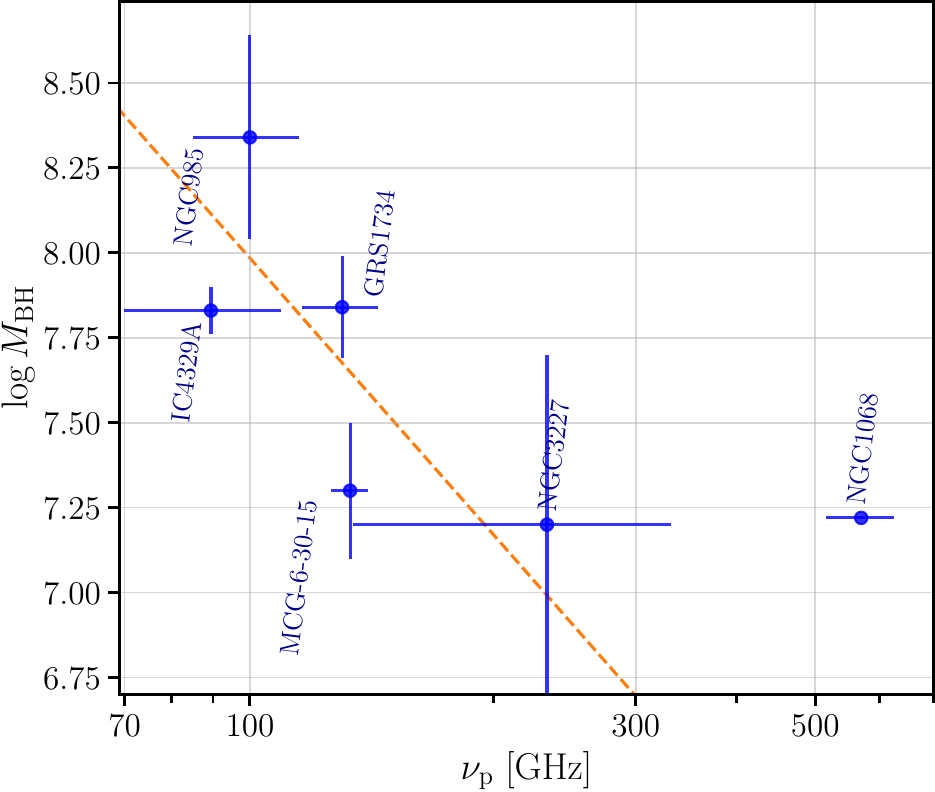} 
    \caption{Peak frequency of the corona component vs. the mass of the SMBH. The dotted lined with slope $-0.37$ shows the model prediction (Sect.~\ref{sec:parameter_exploration}). We note that the errorbar of $M_\mathrm{BH}$ is very small in NGC~1068 \citep{Gallagher2024}.}
    \label{fig:nup_vs_MBH}
\end{figure}

The last thing we study here is whether the peak of the corona SED is anticorrelated with the SMBH mass as predicted in Sect.~\ref{sec:parameter_exploration}. We show a plot of these two quantities in Fig.~\ref{fig:nup_vs_MBH}. Although the sample is still very small, a hint of this anticorrelation can be seen. 

\subsection{$M_\mathrm{BH}$--$L_\mathrm{220\,GHz}$ correlation modelling}

\begin{figure}
    \centering
    \includegraphics[width=\linewidth]{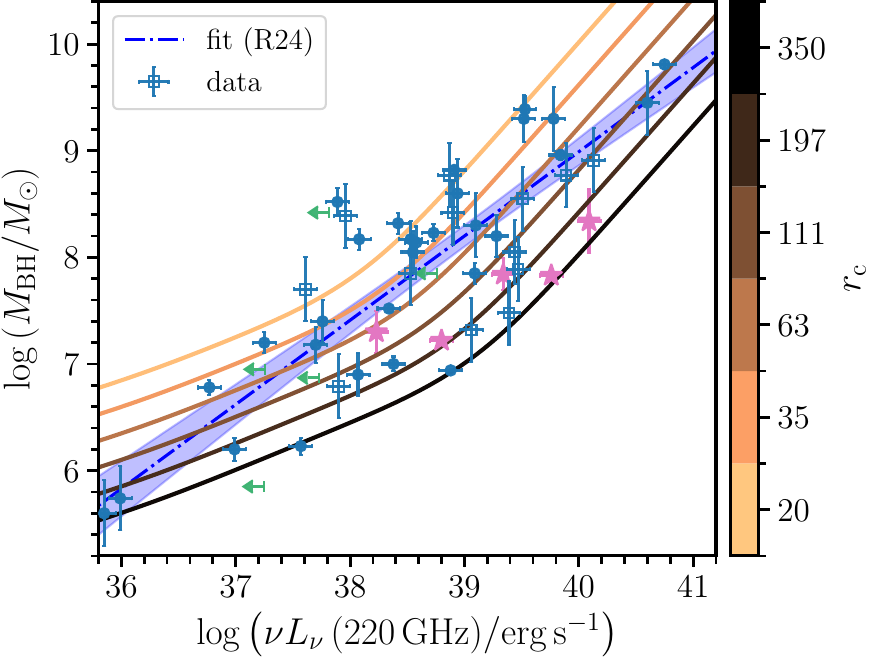} 
    \includegraphics[width=\linewidth]{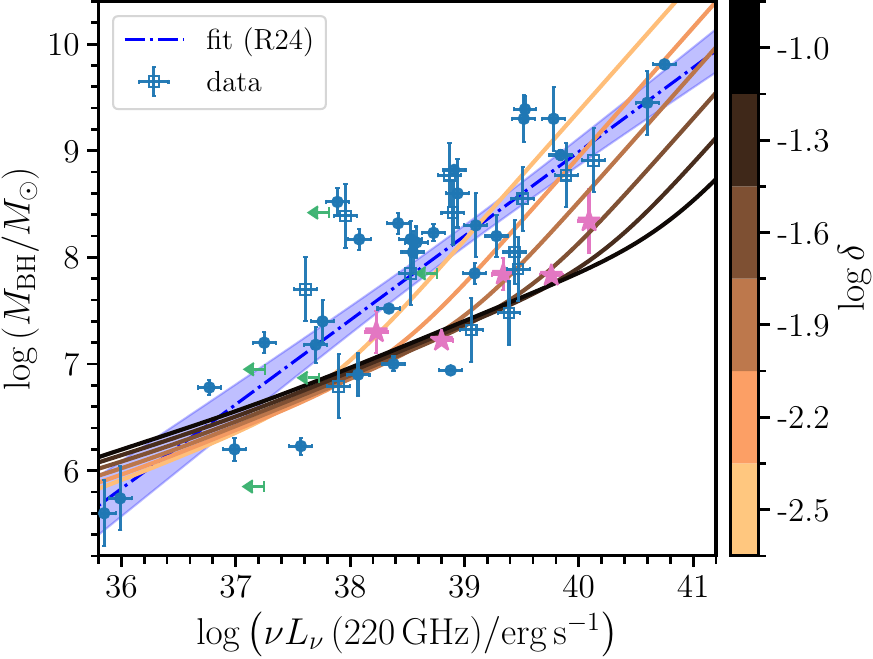} 
    \caption{Correlation between the luminosity at 220~GHz and the black hole mass. The points and the blue curve are from \cite{Ruffa2024}; filled circles are sources with dynamical $M_\mathrm{BH}$ measurements and open squares sources with $M_\mathrm{BH}$ derived using the $M_\mathrm{BH}$--$\sigma_\star$ relation \citep{Ruffa2024}. Upper limits are plotted in green for easier distinction. We also plot with pink stars the sources studied in this work (except NGC~3227, which was already included in their sample). Colour-gradient curves are the corona luminosities calculated for the model parameters defined in Sect.~\ref{sec:parameter_exploration}, and using the approximate mean values of $r_\mathrm{c}=130$ and $\log{\delta}=-2.0$ (Sect.~\ref{sec:results_sources}). We explore different values of $r_\mathrm{c}$ (top panel) and $\delta$ (bottom panel), covering the ranges derived for the sources in the sample.}
    \label{fig:correlation}
\end{figure}

For a fixed value of $M_\mathrm{BH}$, we can use our corona emission model to calculate the corona emission at a rest-frame frequency of 220~GHz. We can repeat this calculation for a broad range of $M_\mathrm{BH}$ and plot this relation as in Fig.~\ref{fig:correlation}. One key prediction of our model is a slope change in the correlation at $\nu L_\nu \sim 10^{39}$\ergs, related to the corona emission becoming optically thin at 220~GHz for SMBHs more massive than $\sim$10$^8$\Msun (Fig.~\ref{fig:corona_sed_M}). 

We compare our correlation with the empirical one reported by \cite{Ruffa2024}. The result is shown in Fig.~\ref{fig:correlation}. In general, we can broadly explain the spread on the observed correlation when taking into account the parameter space available in $\delta$ and $r_\mathrm{c}$, which are the most relevant parameters for the corona emission. This supports a physical connection underlying this correlation, although this still needs to be tested in more detail and with an increased sample size, in particular at $\nu L_\nu \lesssim 10^{37}$\ergs and $\nu L_\nu \gtrsim 10^{40}$\ergs. In addition, when using a single frequency band, very-high-angular-resolution observations are needed to isolate the (compact) coronal component from other galactic contributions, which can otherwise contaminate significantly the flux densities at mm frequencies. This additional non-coronal emission shifts the observed sources farther to the right in Fig.~\ref{fig:correlation}. We also note that some points lie significantly to the left of the correlation (in particular around $\nu L_\nu \sim 10^{38}$\ergs), suggesting that they are underluminous in mm wavelengths. 

\section{Discussion}

The corona properties we derived in Sect.~\ref{sec:results_sources} are roughly in agreement with what has been reported by other authors in the literature \citep{Behar2015, Inoue2018}, which raises two points. First, we note that the sizes inferred from the mm observations are consistently larger than the ones inferred from X-ray observations \citep{Fabian2015} or microlensing \citep{Rybak2025}. Second, we note that the value we obtain of $\delta \approx 1.4$\% is in agreement with the value required to explain the MeV background \citep{Inoue2008, Inoue2019}. In addition, this value of $\delta$ corresponds to $\epsilon_B \sim 0.5$, leading to $\beta_B \sim 10$ (Sect.~\ref{sec:corona_model}). We note that values of $\beta_B > 1$ may challenge the ability of magnetic fields to sustain particle acceleration and heating in the corona \citep{Inoue2024}.

Below, we provide an overview of the main limitations of our multi-component SED fitting, both from the modelling perspective and the observational aspects, in order to suggest possible ways to move forward.

First, we highlight that we are using a one-zone coronal emission model. Even though the assumption of a homogeneous emitter is most likely an oversimplification, it still yields results that are currently consistent with the observations. It remains extremely difficult to constrain more complex models \cite[e.g.][]{Raginski2016} unless the optically thick part of the radio-mm SED is clearly detected. In most sources, this will require very deep and high-angular-resolution observations at frequencies below 100~GHz. Notably, the correlation between radio and X-ray luminosity is weaker at 22~GHz \citep{Magno2025} than at 100~GHz \citep{Ricci2023}, consistent with the weaker coronal emission at 22~GHz predicted by our one-zone model. Another limitation of assuming a homogeneous spherical corona is that we cannot predict realistic images of it, although their characteristic sizes of below a few $\mu$as make them unresolvable with current facilities. As a final note on the coronae, we note that when observing very dust-obscured sources, we could expect absorption of its emission by dust at frequencies above $\nu_{\tau_1}$, although this depends on the dust being along our line of sight to the corona. 

Further limitations of the methodology used here include not accounting for an AME component from spinning dust grains, which could be relevant in sources that present a bump in the SED at $\sim$30~GHz (which is not the case for the sources in our sample). This component could be added at the expense of additional free parameters. Regarding the synchrotron component, we preferred to ignore the effects of f--f absorption (FFA) in this component due to the limited data at high angular resolution at low frequencies, although this possibility is available in our code \citep[an application of a clumpy absorption medium was presented in][]{Mutie2025}. In this regard, it could be possible to adopt a more model-dependent approach in which, under certain assumptions about the diffuse gas distribution, the FFA is linked to the f--f emission, but this should be properly tested on a large sample of galaxies not necessarily limited to RQ AGN with corona emission. For this reason, we currently favour a more phenomenological approach in which the parameters that describe the FFA and f--f emission are not tied to further assumptions.  

Finally, we highlight the observational challenges involved in studying the corona emission of RQ AGN. From a fitting perspective, it is crucial to have a well-sampled SED from radio to FIR with consistent angular resolutions and $uv$-coverage. This ensures that the same scales are probed, although it is very challenging to obtain such a consistent dataset \citep[so far only attempted in][]{Mutie2025}. High-resolution IR imaging is currently unavailable, which hinders our chances of constraining the dust component. The next generation VLA (ngVLA\footnote{\url{https://ngvla.nrao.edu}}) is very promising due to the extremely good angular resolution and sensitivity to frequencies up to $\sim$100 GHz that it will provide. In the mm to sub-mm domain, ALMA will remain the most suitable observatory. 

One of the most difficult issues to address in the SED fitting is the corona variability on timescales of days. So far, the most extreme case of variability was found in \icf, in which the mm flux density varied by a factor $\sim$3 and the inferred corona  size by a factor of $\sim$2, while $\delta$ remained about constant \citep{Shablovinskaya2024}. Other sources have shown even less variability, and thus the corresponding change in the coronal parameters should be even smaller. However, this is difficult to assess due to the very limited studies of this phenomenon to date, and the lack of broad frequency coverage needed to get a good SED fit. 
In this regard, the prospects for the ALMA Wide Sensitivity Upgrade are particularly promising, as it will allow us to obtain a broad bandwidth in a single epoch, something particularly useful around $\sim$100~GHz \citep[as shown in][]{Shablovinskaya2024}. Complementing this with simultaneous hard X-ray observations is ideal, as it can further reduce the uncertainty in the corona parameters $\tau_\mathrm{T}$ and $k\,T_\mathrm{c}$, although this should not be a deciding factor in most cases (Fig.~\ref{fig:corona_seds_tied_parameters}).

\section{Conclusions}

We presented a comprehensive study of mm-wave coronal emission in SMBH coronae. We first introduced a model for calculating the synchrotron emission from an SMBH corona based on \cite{Inoue2018}, coupled with an SED fitting scheme optimised for realistic datasets. That is, we minimised the number of free parameters in the model by introducing a physically motivated parameterization of the magnetic field intensity and an empirical parameterization for the opacity and temperature of the corona. We applied our model systematically to the five RQ AGNs for which coronal mm emission had been reported, together with two additional RQ AGNs with excellent multi-wavelength data from radio-cm to sub-mm frequencies. Our SED fitting results suggest that the coronae have non-thermal particle energy density fractions of 0.5--10\% and radii of 60--250\,$R_\mathrm{g}$, from which we infer typical magnetic fields of 10--60~G. These sets of parameters can also explain the putative correlation between $M_\mathrm{BH}$ and mm-wave luminosity presented in \cite{Ruffa2024}. This opens the possibility of estimating $M_\mathrm{BH}$ via mm-wave observations in systems difficult to observe at other wavelengths, such as extremely obscured sources. Another suggested application of this approach is to measure faint coronal emission in high-$z$ lensed quasars using next-generation radio observatories. We conclude that multifrequency continuum observations spanning from radio to far-IR frequencies---with particular emphasis on the mm band---can be used as direct probes of the physical properties of SMBH coronae. The knowledge gained from these observations can have direct applications to other studies related to accretion disc coronae, such as in XRBs and tidal disruption events. Moreover, the detailed information on electron acceleration in the corona can help multi-messenger emission modelling, allowing to derive more robust predictions of neutrino sources, or verify neutrino detections. 

   
\begin{acknowledgements}
      SdP, CY, SA and SK gratefully acknowledge funding from the European Research Council (ERC) under the European Union's Horizon 2020 research and innovation programme (grant agreement No 789410, PI: S. Aalto).
      SGB acknowledges support from the Spanish grant PID2022-138560NB-I00, funded by MCIN AEI/10.13039/501100011033/FEDER, EU.
      MPS acknowledges support under grants RYC2021-033094-I, CNS2023-145506 and PID2023-146667NB-I00 funded by MCIN/AEI/10.13039/501100011033 and the European Union NextGenerationEU/PRTR.
      JBT acknowledges support from the DFG, via the Collaborative Research Center SFB1491 "Cosmic Interacting Matters - from Source to Signal" (project no.\ 445052434).
      IGB is supported by the Programa Atracci\'on de Talento Investigador ``C\'esar Nombela'' via grant 2023-T1/TEC-29030 funded by the Community of Madrid.

      This paper makes use of the following ALMA data: 
      ADS/JAO.ALMA\#2017.1.00236.S, ADS/JAO.ALMA\#2019.1.00618.S, ADS/JAO.ALMA\#2019.1.01230.S, 
      ADS/JAO.ALMA\#2015.1.01047.S, ADS/JAO.ALMA\#2023.1.00121.S, ADS/JAO.ALMA\#2021.1.00812.S. 
      ALMA is a partnership of ESO (representing its member states), NSF (USA) and NINS (Japan), together with NRC (Canada), NSTC and ASIAA (Taiwan), and KASI (Republic of Korea), in cooperation with the Republic of Chile. The Joint ALMA Observatory is operated by ESO, AUI/NRAO and NAOJ.

      This research has made use of the CIRADA cutout service at \url{cutouts.cirada.ca}, operated by the Canadian Initiative for Radio Astronomy Data Analysis (CIRADA). CIRADA is funded by a grant from the Canada Foundation for Innovation 2017 Innovation Fund (Project 35999), as well as by the Provinces of Ontario, British Columbia, Alberta, Manitoba and Quebec, in collaboration with the National Research Council of Canada, the US National Radio Astronomy Observatory and Australia’s Commonwealth Scientific and Industrial Research Organisation.
      
\end{acknowledgements}

%
%

\bibliographystyle{aa} 
\bibliography{biblio}  

\appendix

\section{Synchrotron emission from a thermal population of electrons} \label{app:thermal_synchrotron}
\begin{figure}
    \centering
    \includegraphics[width=\linewidth]{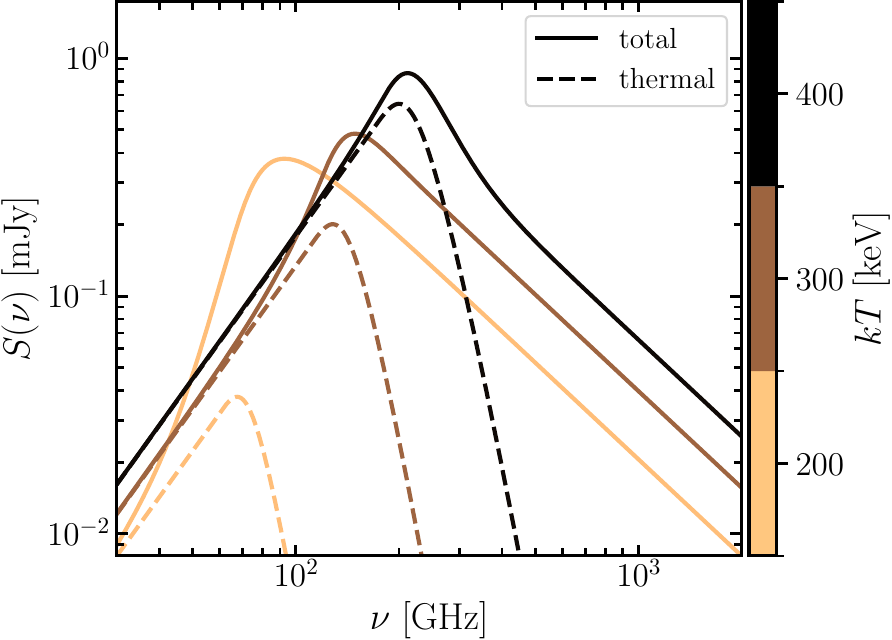} 
    \caption{SEDs for different values of $k T_\mathrm{c}$ for the same parameters as in Sect.~\ref{sec:parameter_exploration} but fixing $\delta=10^{-3}$. Dashed lines represent the emission from thermal electrons, and solid lines represent the total emission.}
    \label{fig:corona_seds_thermal}
\end{figure}

We explore in more detail whether it is feasible for thermal electrons alone to produce appreciable synchrotron emission, compatible with the mm emission detected in several sources. The number of relativistic electrons in a thermal distribution strongly depends on the temperature of the plasma. We, therefore, calculated SEDs for corona temperatures ranging from 200 to 400~keV, shown in Fig.~\ref{fig:corona_seds_thermal}. We note that values of $k T_\mathrm{c} \sim 500$~keV could be unsustainable due to runaway pair production. 
For $k T_\mathrm{c} < 200$~keV, there are simply not enough relativistic electrons to emit detectable synchrotron, and even for $k T_\mathrm{c} \gtrsim 300$~keV this emission is quite faint. For reference, we also show the total emission, including a very small fraction of non-thermal electrons, $\delta = 10^{-3}$. The non-thermal electrons completely dominate the emission for $k T_\mathrm{c} < 300$~keV. The only case in which the emission from thermal electrons could be dominant is for extremely high corona temperatures $k T_\mathrm{c} \sim 400$~keV and low values of $\delta \lesssim 10^{-3}$. However, such a scenario ($k T_\mathrm{c} \sim 400$~keV, $\tau_\mathrm{T}\sim1$, $r_\mathrm{c} \sim 120$, $B \sim 100$~G) is energetically very demanding, still produces faint mm emission, and would most likely overpredict the X-ray luminosity. We thus conclude that scenarios with $k T_\mathrm{c} < 200$~keV and $\delta > 10^{-3}$ are far more promising for explaining the mm emission in RQ AGNs.

\section{Source sample} \label{app:sample}

In the following, we provide a short description of the sources analysed in this work and their SED fitting. The fluxes reported in this work are a combination of archival and proprietary observations. 
As such, the datasets are non-simultaneous, with typical timespans of several years. Among the archives used, we highlight the LoTSS survey with LOFAR \citep{Shimwell2017}, 
the RACS survey with ASKAP \citep{Hale2021}, and the VLASS survey with the VLA \citep{Lacy2020}. We note that all flux density errors presented in the tables include only the statistical errors; for the inclusion of systematic errors, we refer to Sect.~\ref{sec:parameter_fitting}. For flux densities that are not taken from the literature, we also specify the observatory that took the data we used. We used CARTA\footnote{\url{https://cartavis.org/}} to obtain the flux density fitting a Gaussian profile.

For all sources, we also calculate the predicted X-ray luminosities by plugging the luminosity at 100~GHz from our model fitting into the correlation from \cite{Ricci2023} between the 100-GHz luminosity and the 14--150~keV luminosity. We note that, in principle, we could use the mm luminosity coming only from the corona (which should be the one scaling with the X-ray luminosity), but we prefer to keep consistency with how the correlations in \cite{Ricci2023} were derived, in which the total emission at 100~GHz was used, even though this can have a significant contribution from non-coronal components depending on the source. In addition, in some cases, the corona component peaks at frequencies significantly above 100~GHz, and we speculate that in those cases the correlation of \cite{Kawamuro2022} (K22) between the 230-GHz luminosity and the 14--150~keV luminosity can be more accurate.

\subsection{\grs}
This galaxy is located at a distance of 84~Mpc ($z=0.021$) and is a known hard X-ray source \citep[e.g.][]{Tortosa2017}. Moreover, 
\cite{Michiyama2024} found a $\gtrsim$50\% increase in the source 100-GHz flux in timescales of days; this mm flux variability is a strong indication of the origin of the mm flux in the corona \citep[we refer to][for an in-depth discussion of this reasoning]{Shablovinskaya2024}. Unfortunately, the source also presents a weak radio jet on angular scales of a few arcseconds, making it difficult to build a consistent SED, as the presence of diffuse emission leads to an artificially steep spectrum if the lower resolution data at lower frequencies are included. Due to these limitations on the dataset, and the fact that f--f emission cannot dominate at $\sim$100~GHz, we simply opt to remove the f--f component and to fix $\alpha_\mathrm{sy}=0.75$ as in \cite{Michiyama2024}. In addition, the lack of data above 230~GHz prevents us from constraining the dust component, so we fix $\beta=1.78$ and $\nu_{\tau_1}=800$~GHz.
We adopt a value of $M_\mathrm{BH}=10^{7.84}$\Msun \citep{Mejia-Restrepo2022}. 
Furthermore, we fix $\tau_\mathrm{T}=2.98$ and $k T_\mathrm{c} = 12$~keV, as derived by \cite{Tortosa2017}. We fit the flux densities listed in Table~\ref{table:grs_fluxes}.
We obtain $r_\mathrm{c}=112 \pm 14$ ($R_\mathrm{c}\approx0.5$~ld) and $\log{\delta}=-1.74 \pm 0.13$, leading to $B \sim 20$~G and $\sigma \sim 0.011$; the corona peaks at $\nu_\mathrm{p} = 130 \pm 14$~GHz with a peak flux density of $S_\mathrm{p} = 0.91 \pm 0.11$~mJy and has a luminosity of $L_\mathrm{c} \approx 4\times 10^{39}$\ergs. The predicted X-ray luminosity is $L_\mathrm{14-150\,keV} \approx 8\times 10^{43}$\ergs, corresponding to $\lambda_\mathrm{Edd} \sim 0.004$. The cornerplot of the fit is shown in Fig.~\ref{fig:grs_cornerplot}.

In addition, we explore the change in the corona parameters during the epoch of higher flux, when there are only observations at ALMA B3. Thus, we choose to fix $\log{\delta}=-1.74$ as in the previous epoch, and the fit yields $r_\mathrm{c}=178 \pm 9$. This suggests that the size of the corona can be much larger than the one derived for the better-sampled epoch, although it is not possible at this stage to define which value is more representative of the average size of the mm-emitting corona.

\begin{table}
\caption{Flux densities and sizes of the flux extraction regions for \grs. The Type column indicates if the values were used for the fit as observed (yes), considered as upper limits (UL) or lower limits (LL). References: $^{(a)}$\cite{Michiyama2024} and references therein; $^{(b)}$this work; $^{(c)}$\cite{Marti1998}.}
\label{table:grs_fluxes}
\centering
\begin{tabular}{ccccc}
\toprule
$\nu$ & $S_\nu$ & $\theta_1 \times \theta_2$ & Type & Ref \\
(GHz) & (mJy) & (arcsec$^2$) & & \\
\midrule
1.45    & $51.0 \pm 6.0$     & $7.2 \times 7.2$  & UL & $^{(c)}$  \\
4.89    & $13.0 \pm 1.3$     & $1.9 \times 1.9$  & UL & $^{(c)}$  \\
8.49    & $5.6 \pm 0.6$      & $1.2 \times 1.2$  & UL & $^{(c)}$  \\
14.9    & $2.6 \pm 0.3$      & $0.7 \times 0.7$  & yes & $^{(a)}$  \\
22.0    & $3.8 \pm 0.1$      & $0.55 \times 0.55$& UL & VLA$^{(b)}$\\
22.0    & $1.4 \pm 0.1$      & $0.3 \times 0.3$  & LL & VLA$^{(b)}$\\
97.5    & $1.45 \pm 0.07$    & $0.3 \times 0.3$  & yes & $^{(a)}$  \\
145.0   & $1.44 \pm 0.05$    & $0.3 \times 0.3$  & yes & $^{(a)}$  \\
225.0   & $1.20 \pm 0.07$    & $0.3 \times 0.3$  & yes & $^{(a)}$  \\
\bottomrule
\end{tabular}
\end{table}

\subsection{\icf}

This is a very well-studied source both in radio and X-rays \citep[e.g.][]{Inoue2018, Ingram2023, Tortosa2024, Shablovinskaya2024}, showing convincing evidence of a hybrid corona. In fact, this source is one of the first two sources for which a bump in the SED around $\sim$100~GHz was detected \citep{Inoue2018}. Moreover, \cite{Shablovinskaya2024} proved that the emission at 100~GHz is significantly variable on daily timescales. 
We adopt $M_\mathrm{BH} = 6.8\times 10^7$~\Msun as in \cite{Shablovinskaya2024}, and fit our model to the same dataset presented in Table~1 of \cite{Inoue2018}. To present a consistent approach as with the other sources, we fix $p=2.7$ and $\tau_\mathrm{T}=0.25$ and $kT_\mathrm{c} \approx 166$\,K. 
We note that, observationally, the values of $\tau_\mathrm{T}$ and $kT_\mathrm{c}$ are not well constrained and are also variable \citep[e.g.][]{Tortosa2024}, but the specific values adopted should not play a significant role in the fitting (as shown in Sect.~\ref{sec:parameter_exploration}). The SED is poorly constrained between 10 and 100~GHz; we therefore fix $\alpha_\mathrm{sy}=-0.59$ as in \cite{Inoue2018}. We obtain $r_\mathrm{c}=261^{+141}_{-45}$ and $\log{\delta}=-2.25^{+0.24}_{-0.48}$ ($R_\mathrm{c}\approx 1$~ld), leading to $B \sim 9.3$~G and $\sigma \sim 0.06$. The corona component peaks at $\nu_\mathrm{p} = 90\pm20$~GHz, with a peak flux density of $S_\mathrm{p} = 4.2\pm1.1$~mJy, and has a luminosity of $L_\mathrm{c} \approx 7\times 10^{39}$\ergs. The cornerplot of the fit is shown in Fig.~\ref{fig:icf_cornerplot}.
The predicted X-ray luminosity is $L_\mathrm{14-150\,keV} \approx 2\times 10^{44}$\ergs, corresponding to $\lambda_\mathrm{Edd} \sim 0.02$. 
Alternatively, the SED fit to the multiple epochs of 100-GHz observations presented in \cite{Shablovinskaya2024} yielded a broader range of sizes of the corona of $\sim$170--300~$R_\mathrm{g}$.

\begin{table}
\caption{Flux densities and sizes of the flux extraction regions for \icf. The Type column indicates if the values were used for the fit as observed (yes), which are considered upper limits (UL) or lower limits (LL). References: $^{(a)}$\cite{Inoue2018}; $^{(b)}$\cite{Ichikawa2019}; $^{(c)}$this work; $^{(d)}$\citep{Imanishi2016}.}
\label{table:ic4329_fluxes}
\centering
\begin{tabular}{ccccc}
\toprule
$\nu$ & $S_\nu$ & $\theta_1 \times \theta_2$ & Type & Ref \\
(GHz) & (mJy) & (arcsec$^2$) & & \\
\midrule
1.4    & $60.5 \pm 2.4$      & $8.1 \times 3.6$     & UL  & $^{(a)}$ \\
1.5    & $63.5 \pm 4.8$      & $45.0 \times 45.0$   & UL  & $^{(a)}$ \\
4.9    & $19.5 \pm 0.8$      & $1.3 \times 0.5$     & UL  & $^{(a)}$ \\
8.4    & $13.9 \pm 1.2$      & $0.8 \times 0.4$     & yes  & $^{(a)}$ \\
22.0   & $10.1 \pm 0.1$      & $2.65 \times 0.83$   & UL  & VLA$^{(c)}$ \\
33.0   & $14.3 \pm 1.6$      & $8.5 \times 5.0$     & UL  & $^{(a)}$ \\
35.0   & $14.7 \pm 1.7$      & $8.0 \times 4.7$     & UL  & $^{(a)}$ \\
90.5   & $8.00 \pm 0.49$     & $0.45 \times 0.44$   & yes  & $^{(a)}$ \\
92.4   & $7.92 \pm 0.49$     & $0.44 \times 0.43$   & yes  & $^{(a)}$ \\
102.5  & $8.07 \pm 0.50$     & $0.43 \times 0.40$   & yes  & $^{(a)}$ \\
104.5  & $7.81 \pm 0.47$     & $0.42 \times 0.39$   & yes  & $^{(a)}$ \\
138.0  & $6.69 \pm 0.38$     & $0.28 \times 0.25$   & yes  & $^{(a)}$ \\
139.9  & $6.79 \pm 0.39$     & $0.27 \times 0.25$   & yes  & $^{(a)}$ \\
150.0  & $6.45 \pm 0.39$     & $0.29 \times 0.24$   & yes  & $^{(a)}$ \\
152.0  & $6.27 \pm 0.38$     & $0.25 \times 0.23$   & yes  & $^{(a)}$ \\
213.0  & $4.39 \pm 1.21$     & $0.17 \times 0.15$   & yes  & $^{(a)}$ \\
215.0  & $4.85 \pm 0.84$     & $0.18 \times 0.16$   & yes  & $^{(a)}$ \\
229.0  & $5.17 \pm 0.92$     & $0.18 \times 0.14$   & yes  & $^{(a)}$ \\
231.0  & $4.98 \pm 0.94$     & $0.16 \times 0.14$   & yes  & $^{(a)}$ \\
263.0  & $13.0 \pm 0.2$     & $0.95 \times 0.53$    & UL  & $^{(d)}$ \\
1199   & $488 \pm 73$        & $>5.0 \times 5.0$     & UL  & $^{(b)}$ \\
\bottomrule
\end{tabular}
\end{table}

\subsection{\mcgs}

This RQ-AGN is located at a distance of $D=34$~Mpc and hosts an SMBH with $\log{M_\mathrm{BH}} \approx 7.3$ \citep[][and references therein]{Raimundo2013}. From hard X-ray observations it is inferred $kT_\mathrm{c} \approx 63$~keV and $\tau_\mathrm{T}=0.27$.  
An analysis of the VLA fluxes from images at different resolutions indicates that there is significant diffuse emission on scales $>0.3"$ (Table~\ref{table:mcg6_fluxes}).

We cannot constrain the dust component properly due to the lack of data at $\nu > 300$~GHz with comparable resolution.
Following \cite{Kawamuro2022}, we assume that the dust contribution is not significant at mm wavelengths given the very high angular resolution of the observations. We thus fix the dust component normalization to an arbitrary low value. In addition, trying to fit the spectral index $p$ leads to a solution that hits the hard limit on the priors, and we thus fix $p=2.1$. We obtain $r_\mathrm{c}=115\pm6$ ($R_\mathrm{c}\approx 0.13$~ld) and $\log{\delta}=-1.98 \pm 0.06$, leading to $B \sim 20.9$~G and $\sigma \sim 0.036$. The corona component peaks at $\nu_\mathrm{p} = 133\pm6$~GHz, with a peak flux density of $S_\mathrm{p} = 0.64\pm0.06$~mJy, and has a bolometric synchrotron luminosity of $L_\mathrm{c}\approx 3\times10^{38}$\ergs. 
The predicted X-ray luminosity is $L_\mathrm{14-150\,keV} \sim 8\times10^{43}$\ergs, corresponding to an Eddington ratio of $\lambda_\mathrm{Edd} \sim 0.003$. The cornerplot of the fit is shown in Fig.~\ref{fig:mcgs_cornerplot}.

\begin{table}
\caption{Flux densities and sizes of the flux extraction regions for \mcgs. The Type column indicates if the values were used for the fit as observed (yes), considered as upper limits (UL) or lower limits (LL). References: $^{(a)}$this work; $^{(b)}$\cite{Ichikawa2019}.}
\label{table:mcg6_fluxes}
\centering
\begin{tabular}{ccccc}
\toprule
$\nu$ & $S_\nu$ & $\theta_1 \times \theta_2$ & Type & Ref \\
(GHz) & (mJy) & (arcsec$^2$) & & \\
\midrule
1.4      & $1.55 \pm 0.3$    & $10.0 \times 10.0$   & UL  & VLA$^{(a)}$ \\
3.0      & $0.94 \pm 0.2$    & $3.6 \times 2.0$     & UL  & VLA$^{(a)}$ \\
4.9      & $0.9 \pm 0.2$     & $1.4 \times 1.4$     & UL  & VLA$^{(a)}$ \\
5.5      & $0.28 \pm 0.06$   & $1.5 \times 0.7$     & UL  & VLA$^{(a)}$ \\
8.46     & $0.65 \pm 0.06$   & $1.3 \times 0.5$     & UL  & VLA$^{(a)}$ \\
8.49     & $0.12 \pm 0.04$   & $0.52 \times 0.22$   & UL & VLA$^{(a)}$ \\
9.0      & $0.408 \pm 0.035$ & $0.8 \times 0.3$     & UL  & VLA$^{(a)}$ \\
90.52    & $0.59 \pm 0.03$   & $0.090 \times 0.090$ & yes & ALMA$^{(a)}$ \\
92.42    & $0.59 \pm 0.03$   & $0.090 \times 0.090$ & yes & ALMA$^{(a)}$ \\
102.52   & $0.57 \pm 0.03$   & $0.090 \times 0.090$ & yes & ALMA$^{(a)}$ \\
104.48   & $0.60 \pm 0.03$   & $0.090 \times 0.090$ & yes & ALMA$^{(a)}$ \\
226.31   & $0.56 \pm 0.03$   & $0.123 \times 0.123$ & yes & ALMA$^{(a)}$ \\
240.83   & $0.53 \pm 0.03$   & $0.123 \times 0.123$ & yes & ALMA$^{(a)}$ \\
228.81   & $0.59 \pm 0.03$   & $0.123 \times 0.123$ & yes & ALMA$^{(a)}$ \\
243.08   & $0.54 \pm 0.03$   & $0.123 \times 0.123$ & yes & ALMA$^{(a)}$ \\
348.4    & $1.0 \pm 0.2$     & $0.27 \times 0.24$   & UL  & ALMA$^{(a)}$ \\
856.6    & $86 \pm 8$        & $>5.0 \times 5.0$    & UL  & $^{(b)}$ \\
1199     & $233 \pm 14$      & $>5.0 \times 5.0$    & UL  & $^{(b)}$ \\
\bottomrule
\end{tabular}
\end{table}

\subsection{\mcge}

\begin{figure}
    \centering
    \includegraphics[width=\linewidth]{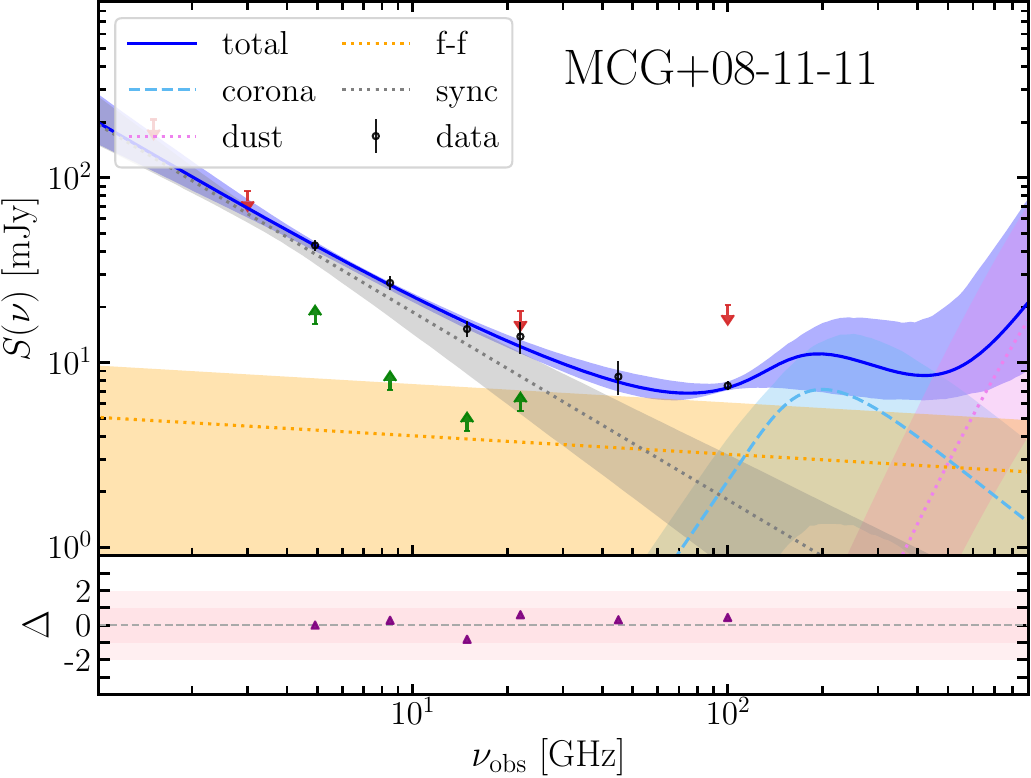} 
    \caption{Same as Fig.~\ref{fig:sources_seds_non-cons} but for \mcge. The corona component is not inferred from the SED shape, but rather from the variability \citep{Petrucci2023}.}
    \label{fig:SED_mgce}
\end{figure}

This LIRG source is located at $z=0.02$ and hosts an SMBH with a mass of $\log{M_\mathrm{BH}} = 7.81$ \citep{Koss2017}. This is a well-known hard X-ray source, and \cite{Petrucci2023} showed evidence of a correlated variability in the hard X-ray and mm emission at 100~GHz (using NOEMA observations).
In \cite{Smith2020}, it is listed as a jetted source at 22~GHz. 
VLA images at 8.5 and 14.9~GHz show that the inner structure has three point-like sources within 0.5", plus diffuse emission on larger scales. For this reason, we integrate the flux within a $\approx$1" region at all frequencies. Morevoer, \cite{Pal2023c} measured $kT_\mathrm{c} \sim $30--60~keV, suggesting that the source has $\tau_\mathrm{T} \gtrsim 1$; we fix $\tau_\mathrm{T} = 1.5$. 
The fact that the $\sim$100-GHz emission reported by \cite{Petrucci2023} is variable on hour-timescales indicates that the corona contribution is non-negligible at this frequency.
CARMA observations at 100~GHz with a 1"-resolution indicate a flux density of 7.5~mJy \citep{Behar2018}, which is significantly lower than the flux density of 18~mJy reported by \citep{Petrucci2023}. 
We fixed the dust component parameters $\beta=1.78$ and $\nu_{\tau_1}=800$~GHz.
The corona SED is, unfortunately, very poorly constrained, and a robust fit cannot be addressed at this stage without high-resolution observations at frequencies $>$100~GHz. Nonetheless, we can at least confirm that the observations are consistent with a corona of $r_\mathrm{c}\sim260$ ($R_\mathrm{c}\sim1$~ld) and $\log{\delta} \sim-1.4$ (yielding $B\sim30$~G).  
The predicted X-ray luminosity is $L_\mathrm{14-150\,keV} \sim 3\times10^{44}$\ergs, corresponding to an Eddington ratio of $\lambda_\mathrm{Edd} \sim 0.035$. The SED is shown in Fig.~\ref{fig:SED_mgce} and the cornerplot of the fit is shown in Fig.~\ref{fig:mcge_cornerplot}. 

\begin{table}
\caption{Flux densities and sizes of the flux extraction regions for \mcge. The Type column indicates if the values were used for the fit as observed (yes), considered as upper limits (UL) or lower limits (LL). References: $^{(a)}$this work; $^{(b)}$\cite{Smith2020}; $^{(c)}$\cite{Panessa2022}; $^{(d)}$\cite{Behar2018}; $^{(e)}$\cite{Petrucci2023}; $^{(f)}$\cite{Ichikawa2019}.}
\label{table:mcg8_fluxes}
\centering
\begin{tabular}{ccccc}
\toprule
$\nu$ & $S_\nu$ & $\theta_1 \times \theta_2$ & Type & Ref \\
(GHz) & (mJy) & (arcsec$^2$) & & \\
\midrule
1.51     & $197 \pm 1$      & $28.4 \times 28.4$  & UL & VLA$^{(a)}$ \\
3.0      & $74 \pm 10$      & $2.6 \times 3.0$    & UL & VLA$^{(a)}$ \\
4.91     & $17.0 \pm 0.2$   & $0.84 \times 0.84$  & LL & VLA$^{(a)}$ \\
4.91     & $43.1 \pm 2.0$   & $1.1 \times 1.3$    & yes & VLA$^{(a)}$ \\
8.49     & $7.5 \pm 0.1$    & $0.38 \times 0.38$  & LL & VLA$^{(a)}$ \\
8.49     & $27 \pm 2.0$     & $1.1 \times 1.1$    & yes & VLA$^{(a)}$ \\
14.9     & $4.5 \pm 0.1$    & $0.22 \times 0.22$  & LL & VLA$^{(a)}$ \\
14.9     & $15.2 \pm 1.3$   & $1.1 \times 1.1$    & yes & VLA$^{(a)}$ \\
22.0     & $5.95 \pm 0.4$   & $0.14 \times 0.14$  & LL & VLA$^{(a)}$ \\
22.0     & $13.84 \pm 2.6$  & $1.0 \times 1.0$    & yes & $^{(b)}$ \\
22.0     & $15.85 \pm 3.1$  & $6.0 \times 6.0$    & UL & $^{(b)}$ \\
45.0     & $8.4 \pm 1.7$    & $1.0 \times 1.0$    & yes & $^{(b)}$, $^{(c)}$ \\
100.0    & $7.5 \pm 0.2$    & $1.0 \times 1.0$    & yes & $^{(d)}$ \\
100.0    & $18.33 \pm 2.0$  & $>1.0 \times 1.0$   & UL & $^{(e)}$ \\
1199     & $2531 \pm 270$   & $>5.0 \times 5.0$   & UL  & $^{(f)}$ \\
\bottomrule
\end{tabular}
\end{table}

\subsection{\ngcn}
This galaxy located at $D\approx 190$~Mpc hosts an SMBH with $M_\mathrm{BH} = 2.2\times 10^8$\Msun \citep[][and references therein]{Inoue2018}. \cite{Inoue2018} reported an excess in the SED of this source at $\sim$100~GHz from which they inferred the presence of a corona.  
The dust component is essentially unconstrained due to a very bright FIR flux from the whole galaxy \citep{Ichikawa2019}. The high-resolution ALMA data favours a corona with a harder electron population, consistent with $p\approx2.1$ \citep{Inoue2018}. Trying to fit this parameter hits the lower limit in the priors, and thus we fix it to $p=2.1$. Following \cite{Inoue2018}, we also fix $kT_\mathrm{c}=29$~keV and $\tau_\mathrm{T}=3.5$. 
The inferred properties of the corona are: $r_\mathrm{c}=111^{+33}_{-14}$ ($R_\mathrm{c}\approx 1.4$~ld) and $\log{\delta}=-2.03^{+0.13}_{-0.22}$, leading to $B \sim 14.4$~G and $\sigma \sim 0.014$. The corona SED component peaks at $\nu_\mathrm{p} = 100\pm15$~GHz, with a peak flux density of $S_\mathrm{p} = 1.4\pm0.2$~mJy, and has a bolometric synchrotron luminosity of $L_\mathrm{c} \approx 4\times10^{40}$\ergs. We estimate a hard X-ray luminosity $L_\mathrm{14-150\,keV}\approx 4\times 10^{45}$\ergs, corresponding to an Eddington ratio of $\lambda_\mathrm{Edd} \sim 0.011$.
The cornerplot of the fit is shown in Fig.~\ref{fig:ngc985_cornerplot} and the data used is listed in Table~\ref{table:ngcn_fluxes}. 

\begin{table}
\caption{Flux densities and sizes of the flux extraction regions for \ngcn. The Type column indicates if the values were used for the fit as observed (yes), considered as upper limits (UL) or lower limits (LL). References: $^{(a)}$this work; $^{(b)}$\cite{Inoue2018}; $^{(c)}$\cite{Ichikawa2019}.}
\label{table:ngcn_fluxes}
\centering
\begin{tabular}{ccccc}
\toprule
$\nu$ & $S_\nu$ & $\theta_1 \times \theta_2$ & Type & Ref \\
(GHz) & (mJy) & (arcsec$^2$) & & \\
\midrule
1.4      & $5.4 \pm 0.5$    & $6.25 \times 6.25$  & UL  & VLA$^{(a)}$ \\
1.5      & $4.7 \pm 1.4$    & $28.4 \times 28.4$  & UL  & VLA$^{(a)}$ \\
4.9      & $1.2 \pm 0.3$    & $0.5 \times 0.4$    & UL & $^{(b)}$ \\
8.4      & $1.1 \pm 0.2$    & $0.9 \times 0.7$    & UL  & $^{(b)}$ \\
8.49     & $0.89 \pm 0.08$  & $0.26 \times 0.21$  & yes & VLA$^{(a)}$ \\
14.9     & $<0.75$          & $0.40 \times 0.17$  & UL  & VLA$^{(a)}$ \\
19.0     & $1.7 \pm 0.7$    & $16.1 \times 8.6$   & UL  & $^{(b)}$ \\
21.0     & $1.5 \pm 0.7$    & $14.3 \times 7.8$   & UL  & $^{(b)}$ \\
22.5     & $1.2 \pm 0.4$    & $0.39 \times 0.23$  & UL & VLA$^{(a)}$ \\
22.5     & $1.9 \pm 0.3$    & $3.8 \times 1.7$    & UL  & $^{(b)}$ \\
43.3     & $3.1 \pm 1.2$    & $0.3 \times 0.1$    & yes & $^{(b)}$ \\
90.5     & $1.72 \pm 0.12$  & $0.25 \times 0.22$  & yes & $^{(b)}$ \\
92.4     & $1.92 \pm 0.13$  & $0.25 \times 0.21$  & yes & $^{(b)}$ \\
102.5    & $1.64 \pm 0.13$  & $0.23 \times 0.19$  & yes & $^{(b)}$ \\
104.5    & $1.59 \pm 0.13$  & $0.22 \times 0.19$  & yes & $^{(b)}$ \\
138.0    & $1.71 \pm 0.14$  & $0.17 \times 0.14$  & yes & $^{(b)}$ \\
139.9    & $1.54 \pm 0.12$  & $0.18 \times 0.15$  & yes & $^{(b)}$ \\
150.0    & $1.44 \pm 0.12$  & $0.16 \times 0.13$  & yes & $^{(b)}$ \\
151.8    & $1.57 \pm 0.12$  & $0.16 \times 0.13$  & yes & $^{(b)}$ \\
213.0    & $1.35 \pm 0.10$  & $0.12 \times 0.10$  & yes & $^{(b)}$ \\
215.0    & $1.27 \pm 0.10$  & $0.12 \times 0.09$  & yes & $^{(b)}$ \\
229.0    & $1.26 \pm 0.10$  & $0.11 \times 0.09$  & yes & $^{(b)}$ \\
231.0    & $1.22 \pm 0.09$  & $0.11 \times 0.09$  & yes & $^{(b)}$ \\
599.6    & $184 \pm 25$     & $7 \times 7$        & UL  & $^{(c)}$ \\
856.6    & $486 \pm 51$     & $7 \times 7$        & UL  & $^{(c)}$ \\
1199     & $951 \pm 67$     & $7 \times 7$        & UL  & $^{(c)}$ \\
\bottomrule
\end{tabular}
\end{table}

\subsection{NGC~1068}
This galaxy is located at $D \approx 14.6$~Mpc ($z=0.0033$) and hosts an SMBH with $M_\mathrm{BH} = (1.66\pm0.01)\times10^7$\Msun \citep{Gallimore2024}. This source has been the subject of very extensive research, given that it was the first RQ galaxy to be identified as a neutrino source \citep{IceCube2022}. It has also been extensively studied in X-rays, revealing the presence of a corona \citep[e.g.][]{Bauer2015}. Furthermore, \cite{Inoue2020} modelled the radio continuum data and concluded that the synchrotron emission from the corona was responsible for the sub-mm emission detected by ALMA, which means that the corona peaks at a rather high frequency in this source compared to others. In a more recent work, \cite{Mutie2025} presented a consistent SED of the source matched in resolution and \textit{uv}-coverage, confirming the presence of corona emission. For completeness, we reproduce the fluxes from their work in Table~\ref{table:ngc1068_fluxes}.

The nuclear region of this source is particularly complex, and so is its SED at low frequencies \citep{Mutie2025}.
For this reason, here we consider only high-resolution observations at $\nu \geq 15$~GHz. Due to the lack of FIR data to constrain the dust component, we fix $\beta=1.78$ and $\nu_{\tau_1}=800$~GHz as with the other sources. We refer to \cite{Garcia-Burillo2016} for a discussion of the dust emission, although we caution that these authors did not take into account the coronal emission in their analysis. In addition, we fix $\alpha_\mathrm{sy}=-0.5$ as otherwise the spectral index hits the hard limit set on $-0.5$. We obtain $r_\mathrm{c}=67.5 \pm 5.5$ ($R_\mathrm{c}\approx 0.065$~ld) and $\log{\delta}=-0.98 \pm 0.13$, leading to $B \sim 158$~G and $\sigma \sim 1.1$. The corona SED peaks at $\nu_\mathrm{p} \simeq 568 \pm 55$~GHz, with a peak flux density of $S_\mathrm{p} = 10.1\pm1.9$~mJy, and has a bolometric synchrotron luminosity of $L_\mathrm{c} \approx 4.4\times10^{39}$\ergs. Given that the peak frequency is much greater than 100~GHz, we estimate the hard X-ray luminosity using the relation from \cite{Kawamuro2022}; we obtain a luminosity $L_\mathrm{14-150\,keV}\approx 2.3\times 10^{43}$\ergs, corresponding to an Eddington ratio of $\lambda_\mathrm{Edd} \sim 0.011$.
The cornerplot of the fit is shown in Fig.~\ref{fig:ngc1068_cornerplot}. We note that this fit is quite similar to the one obtained by \cite{Inoue2020} and \cite{Mutie2025}. 

\begin{table}
\caption{Flux densities and sizes of the flux extraction regions for NGC~1068. The Type column indicates if the values were used for the fit as observed (yes), considered as upper limits (UL) or lower limits (LL). References: $^{(a)}$\cite{Mutie2025}.}
\label{table:ngc1068_fluxes}
\centering
\begin{tabular}{ccccc}
\toprule
$\nu$ & $S_\nu$ & $\theta_1 \times \theta_2$ & Type & Ref \\
(GHz) & (mJy) & (arcsec$^2$) & & \\
\midrule
15     & $17 \pm 1$               & $0.1 \times 0.1$      & UL  & $^{(a)}$ \\
18.9   & $13 \pm 0.1$             & $0.06 \times 0.06$    & yes & $^{(a)}$ \\
19.4   & $12.6 \pm 0.1$           & $0.06 \times 0.06$    & yes & $^{(a)}$ \\
19.9   & $12.5 \pm 0.1$           & $0.06 \times 0.06$    & yes & $^{(a)}$ \\
21.9   & $12.1 \pm 0.1$           & $0.06 \times 0.06$    & yes & $^{(a)}$ \\
22.3   & $12.5 \pm 0.1$           & $0.06 \times 0.06$    & yes & $^{(a)}$ \\
22.9   & $12.0 \pm 0.1$           & $0.06 \times 0.06$    & yes & $^{(a)}$ \\         
94     & $11.0 \pm 0.6$           & $0.06 \times 0.06$    & yes & $^{(a)}$ \\         
100    & $10.2 \pm 0.5$           & $0.06 \times 0.06$    & yes & $^{(a)}$ \\   
241    & $10.6 \pm 0.6$           & $0.06 \times 0.06$    & yes & $^{(a)}$ \\   
256    & $9.9 \pm 0.1$            & $0.06 \times 0.06$    & yes & $^{(a)}$ \\   
345    & $16.1 \pm 0.3$           & $0.06 \times 0.06$    & yes & $^{(a)}$ \\   
357    & $13.9 \pm 0.1$           & $0.06 \times 0.06$    & yes & $^{(a)}$ \\   
477    & $18.7 \pm 0.5$           & $0.06 \times 0.06$    & yes & $^{(a)}$ \\   
483    & $19.6 \pm 0.4$           & $0.06 \times 0.06$    & yes & $^{(a)}$ \\   
688    & $20.0 \pm 2.5$           & $0.06 \times 0.06$    & yes & $^{(a)}$ \\   
697    & $17.0 \pm 1.7$           & $0.06 \times 0.06$    & yes & $^{(a)}$ \\   
706    & $16.5 \pm 1.9$           & $0.06 \times 0.06$    & yes & $^{(a)}$ \\   
\bottomrule
\end{tabular}
\end{table}

\subsection{NGC~3227}
This galaxy located at $D\approx 23$~Mpc \citep{Ricci2023} hosts a $M_\mathrm{BH} = 4.2\times 10^7$\Msun SMBH \citep{Behar2015}.
\citet{Behar2015} showed that this source has a significant excess at $\sim$100 GHz. \citet{Sani2012} reported its flux density at 3~mm as observed by the Plateau de Bure with a $\sim$1" resolution. 
In addition, flux densities from ALMA observations at B6 and B7 were reported in \cite{Alonso-Herrero2019}. 
According to the very high angular resolution observations at B3 in \cite{Ricci2023}, the flux from the core region is $\sim$0.85~mJy.
The dust component is poorly constrained due to a very bright FIR flux from the whole galaxy \citep{Ciesla2012}. However, \cite{Alonso-Herrero2019} fitted the IR SED with a dust model that has a flux of $\sim$10~mJy at $10^3$~GHz (their Fig.~4), so that we can at least check that our model fitting is consistent with that derived independently using dust modelling of mid-IR low-resolution observations. 
Unfortunately, it is not possible to fit a dataset with very high angular resolution ($<0.1"$) due to the limited number of data points. We thus use as reference points with resolutions $\sim$0.4"--1", which is not optimal for detecting the relatively faint corona emission. Thus, the properties of the corona are only loosely constrained as $r_\mathrm{c}=67^{+48}_{-26}$ ($R_\mathrm{c}\approx 0.06$~ld) and $\log{\delta}=-2.07^{+0.56}_{-0.48}$, leading to $B \sim 46$~G and $\sigma \sim 0.09$. The corona SED component peak is very poorly constrained at $\nu_\mathrm{p} = 233 \pm 91$~GHz, with a peak flux density of $S_\mathrm{p} = 0.94\pm0.48$~mJy, and has a bolometric synchrotron luminosity of $L_\mathrm{c} \approx 3\times10^{38}$\ergs. Interestingly, the source is undetected in Q band with the VLA in A array configuration (project code AG568), giving an UL to the compact emission produced at 43.3~GHz, which is consistent with the SED fit of the corona component. We estimate a hard X-ray luminosity $L_\mathrm{14-150\,keV}\approx 8\times 10^{42}$\ergs, corresponding to an Eddington ratio of $\lambda_\mathrm{Edd} \sim 0.004$.
The cornerplot of the fit is shown in Fig.~\ref{fig:ngc3227_cornerplot}.

\begin{table}
\caption{Flux densities and sizes of the flux extraction regions for NGC~3227. The Type column indicates if the values were used for the fit as observed (yes), considered as upper limits (UL) or lower limits (LL). References: $^{(a)}$\cite{Behar2015}; $^{(b)}$\cite{Alonso-Herrero2019}; $^{(c)}$this work; $^{(d)}$\cite{Sani2012}; $^{(e)}$\cite{Bontempi2012}; $^{(f)}$\cite{Smith2020}; $^{(g)}$\cite{Sargent2025}; $^{(h)}$\cite{Ciesla2012}.
}
\label{table:ngc3227_fluxes}
\centering
\begin{tabular}{ccccc}
\toprule
$\nu$ & $S_\nu$ & $\theta_1 \times \theta_2$ & Type & Ref \\
(GHz) & (mJy) & (arcsec$^2$) & & \\
\midrule
0.887   & $117.0 \pm 15.0$  & $25 \times 25$        & UL & RACS$^{(c)}$ \\
1.4     & $82.7 \pm 0.2$    & $5.7 \times 5.7$      & UL & $^{(a)}$ \\
1.4     & $78.2 \pm 5.0$    & $0.3 \times 0.3$      & yes & $^{(a)}$ \\
1.7     & $1.22 \pm 0.12$   & $0.0059 \times 0.0059$ & LL & $^{(e)}$ \\
3.0     & $41.4 \pm 0.5$    & $2.71 \times 2.67$     & UL & VLA$^{(c)}$ \\
5.0     & $0.6 \pm 0.07$    & $0.0012 \times 0.0012$ & LL & $^{(e)}$ \\
5.0     & $15.0 \pm 2.0$    & $0.33 \times 0.33$    & yes & $^{(a)}$ \\
5.0     & $35.0 \pm 0.7$    & $15 \times 15$        & UL & $^{(a)}$ \\
6.0     & $7.89 \pm 0.44$   & $0.30 \times 0.23$    & LL & $^{(g)}$ \\
8.46    & $15.4 \pm 0.5$    & $2.6 \times 2.6$      & UL & VLA$^{(c)}$ \\
8.5     & $12.2 \pm 1.3$    & $0.2 \times 0.2$      & LL & $^{(a)}$ \\
10.0    & $3.75 \pm 0.20$   & $0.18 \times 0.16$    & LL & $^{(g)}$ \\
14.9    & $6.5 \pm 1.1$     & $0.89 \times 0.56$    & yes & VLA$^{(c)}$ \\
15      & $4.7 \pm 0.3$     & $0.13 \times 0.13$    & LL & $^{(a)}$ \\
22      & $6.95 \pm 1.4$    & $6.0 \times 6.0$      & UL & $^{(f)}$ \\
22      & $4.13 \pm 0.8$    & $1.0 \times 1.0$      & UL & $^{(f)}$ \\
43.34   & $<0.121$          & $0.043 \times 0.028$  & no & VLA$^{(c)}$ \\
89      & $1.79 \pm 0.12$   & $1.15 \times 1.05$    & yes & $^{(d)}$ \\
95      & $4.1 \pm 0.2$     & $2.2 \times 2.2$      & UL & $^{(b)}$ \\
95.2    & $2.1 \pm 0.7$     & $0.66 \times 0.45$    & UL & ALMA$^{(c)}$ \\
97.5    & $1.28 \pm 0.35$   & $0.13 \times 0.09$    & LL & ALMA$^{(c)}$ \\
107.9   & $1.27 \pm 0.25$   & $0.16 \times 0.11$    & LL & ALMA$^{(c)}$ \\
230     & $0.7 \pm 0.1$     & $0.2 \times 0.2$      & LL & $^{(b)}$ \\
230     & $1.7 \pm 0.2$     & $0.5 \times 0.5$      & yes & $^{(b)}$ \\
350     & $0.9 \pm 0.1$     & $0.2 \times 0.2$      & LL & $^{(b)}$ \\
350     & $2.0 \pm 0.2$     & $0.5 \times 0.5$      & yes & $^{(b)}$ \\
485     & $1.49 \pm 0.64$   & $0.16 \times 0.12$    & LL & ALMA$^{(c)}$ \\
600     & $1815 \pm 68$     & $>10 \times 10$       & UL & $^{(h)}$ \\
\bottomrule
\end{tabular}
\end{table}

\section{Parameter space and cornerplots} \label{app:parameter_space}
We sample a broad range in the parameter space to ensure that all possible solutions are explored. A list of parameters and their allowed ranges is presented in Table~\ref{table:parameters}. We highlight that we fit the logarithm of those parameter that can spread over several orders of magnitude (namely the normalisations and $\delta$).

\begin{figure*}
    \centering
    \includegraphics[width=\linewidth]{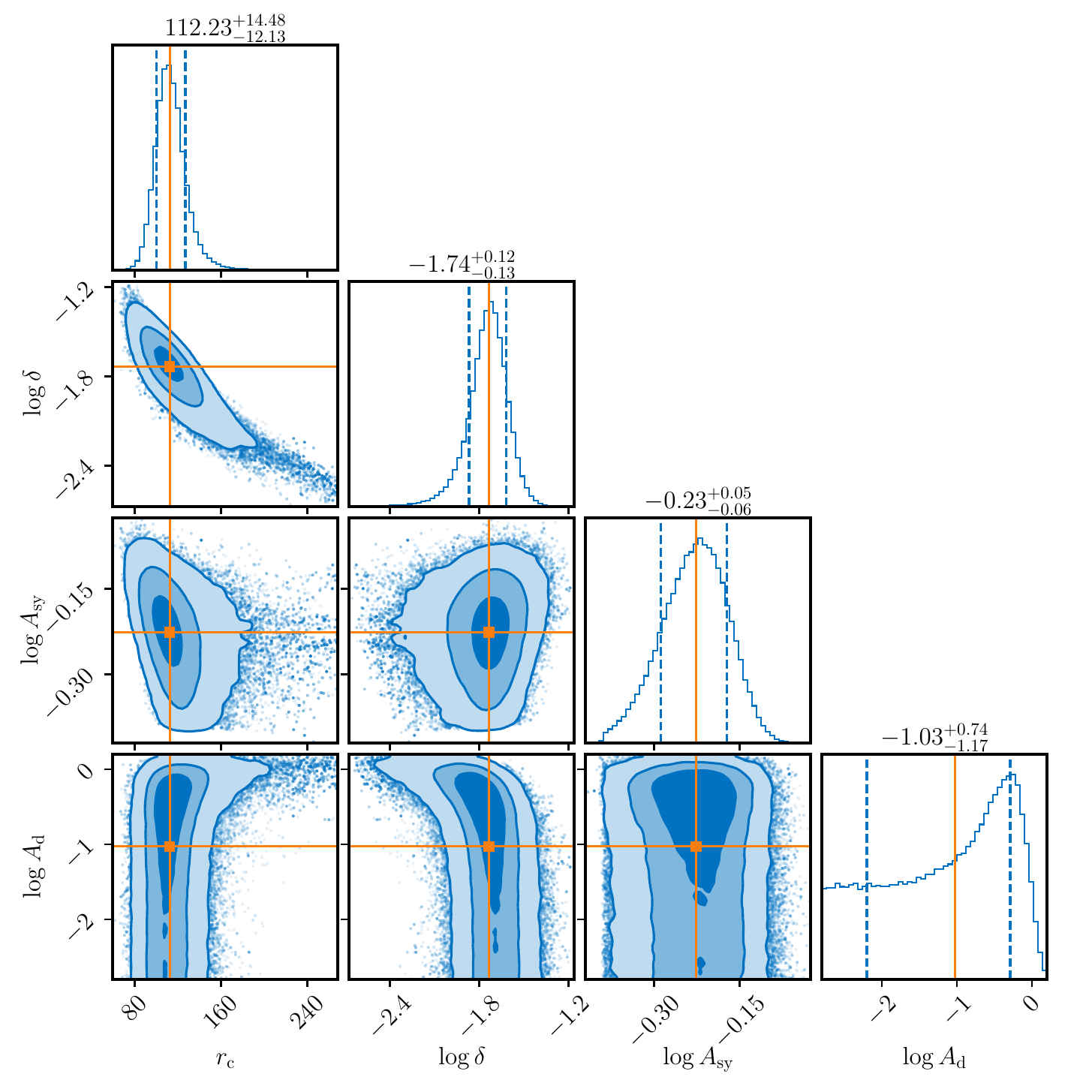} 
    \caption{Cornerplot for the posteriors of the SED fit for \grs.}
    \label{fig:grs_cornerplot}
\end{figure*}

\begin{figure*}
    \centering
    \includegraphics[width=\linewidth]{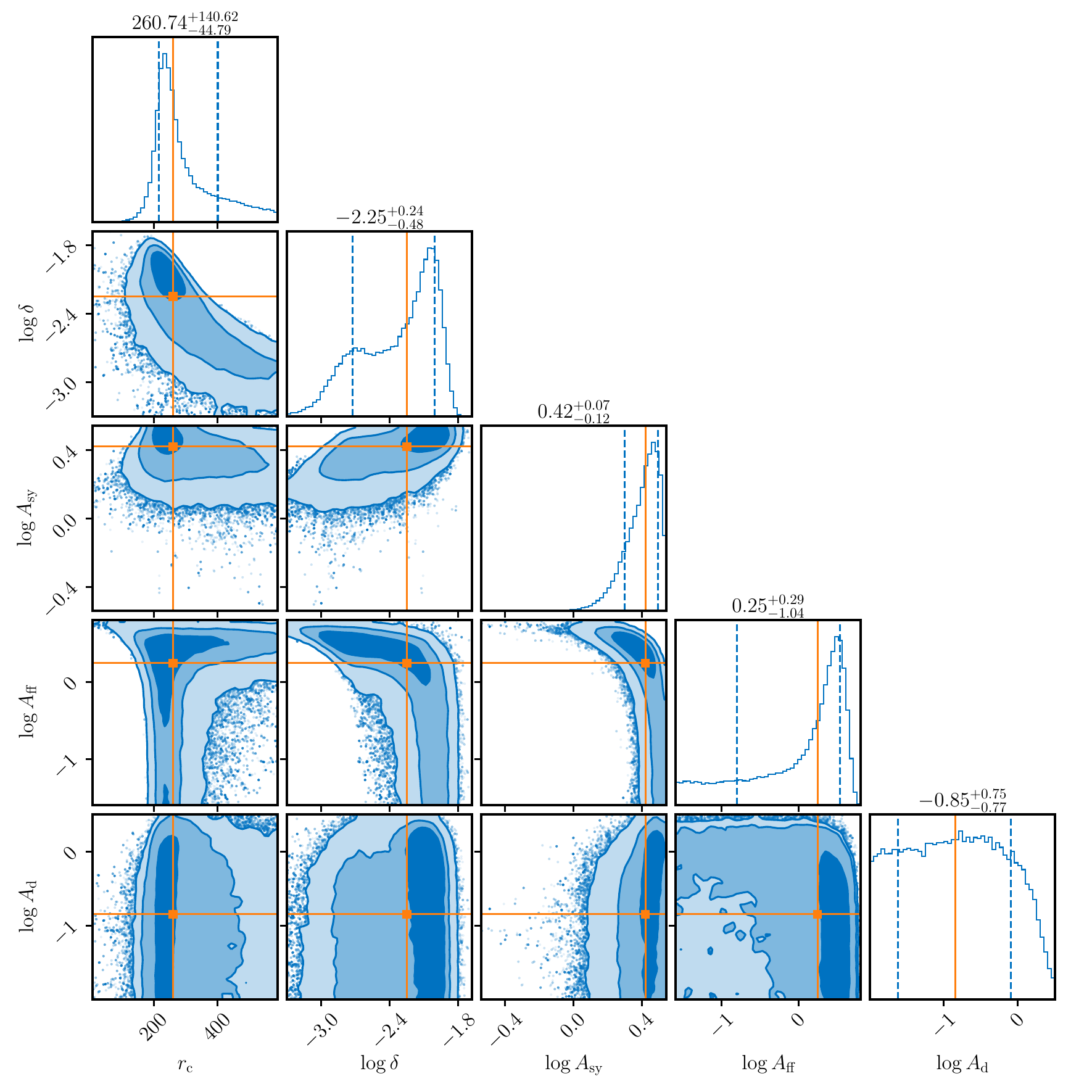} 
    \caption{Cornerplot for the posteriors of the SED fit for \icf.}
    \label{fig:icf_cornerplot}
\end{figure*}

\begin{figure*}
    \centering
    \includegraphics[width=\linewidth]{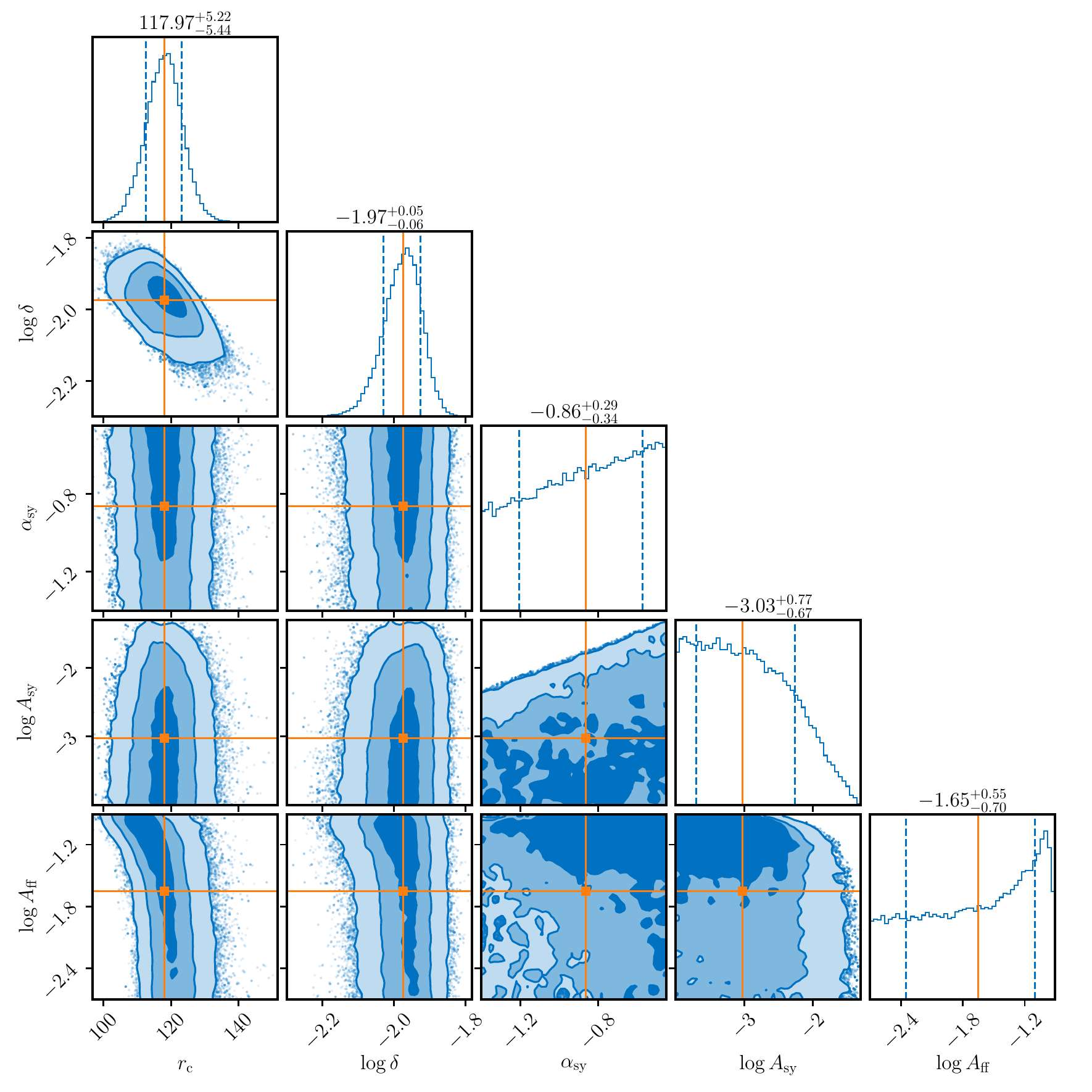} 
    \caption{Cornerplot for the posteriors of the SED fit for \mcgs.}
    \label{fig:mcgs_cornerplot}
\end{figure*}

\begin{figure*}
    \centering
    \includegraphics[width=\linewidth]{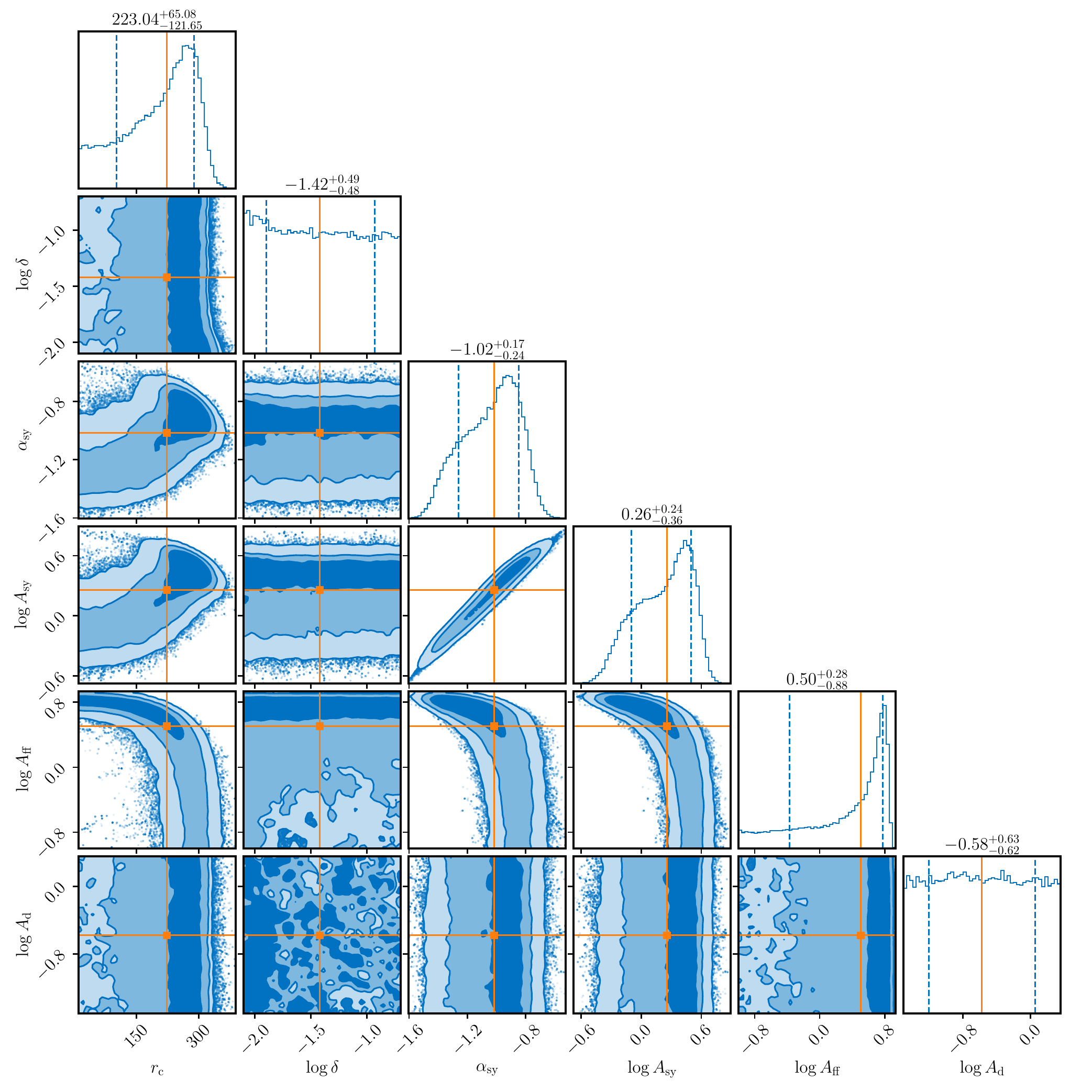} 
    \caption{Cornerplot for the posteriors of the SED fit for \mcge.}
    \label{fig:mcge_cornerplot}
\end{figure*}

\begin{figure*}
    \centering
    \includegraphics[width=\linewidth]{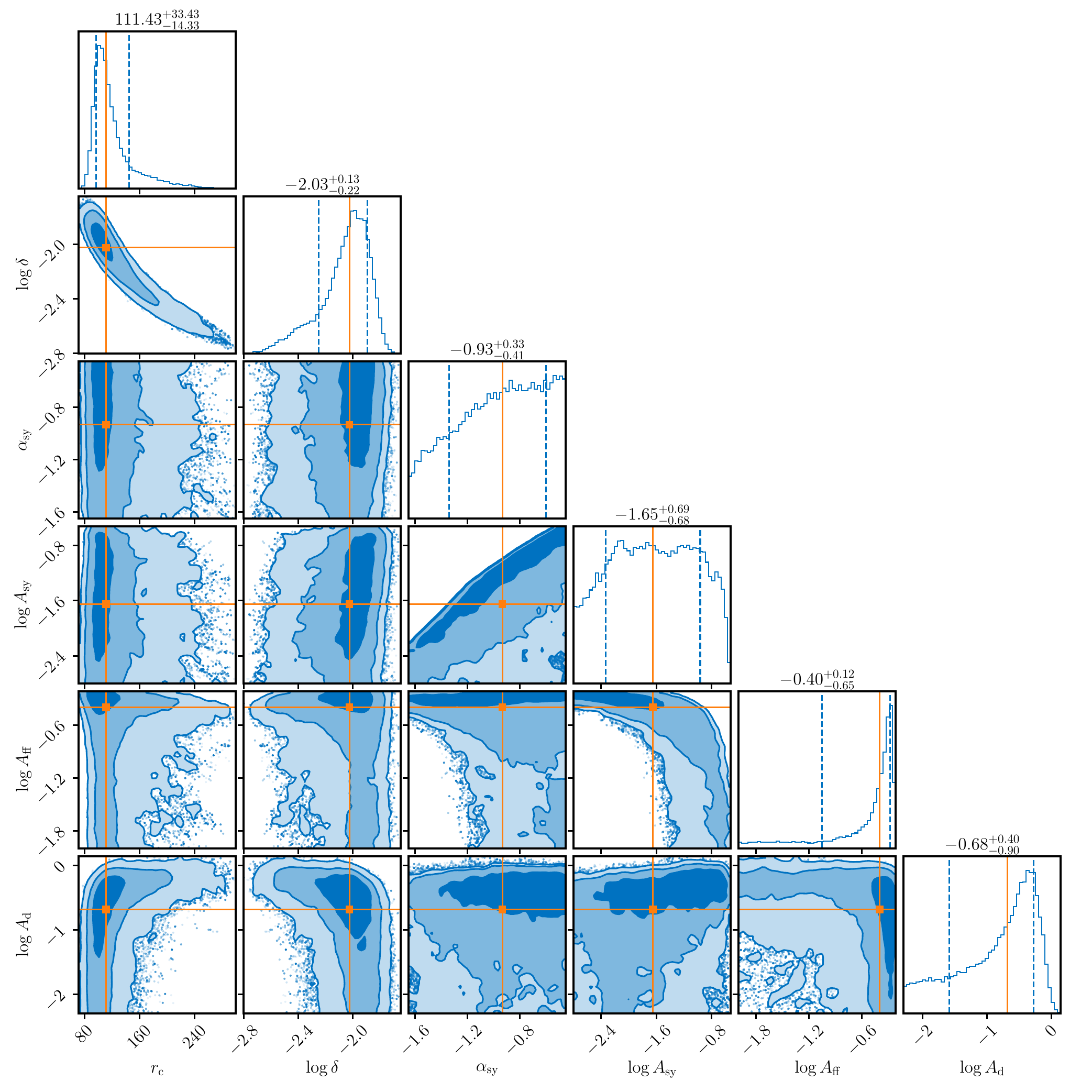} 
    \caption{Cornerplot for the posteriors of the SED fit for \ngcn.}
    \label{fig:ngc985_cornerplot}
\end{figure*}

\begin{figure*}
    \centering
    \includegraphics[width=\linewidth]{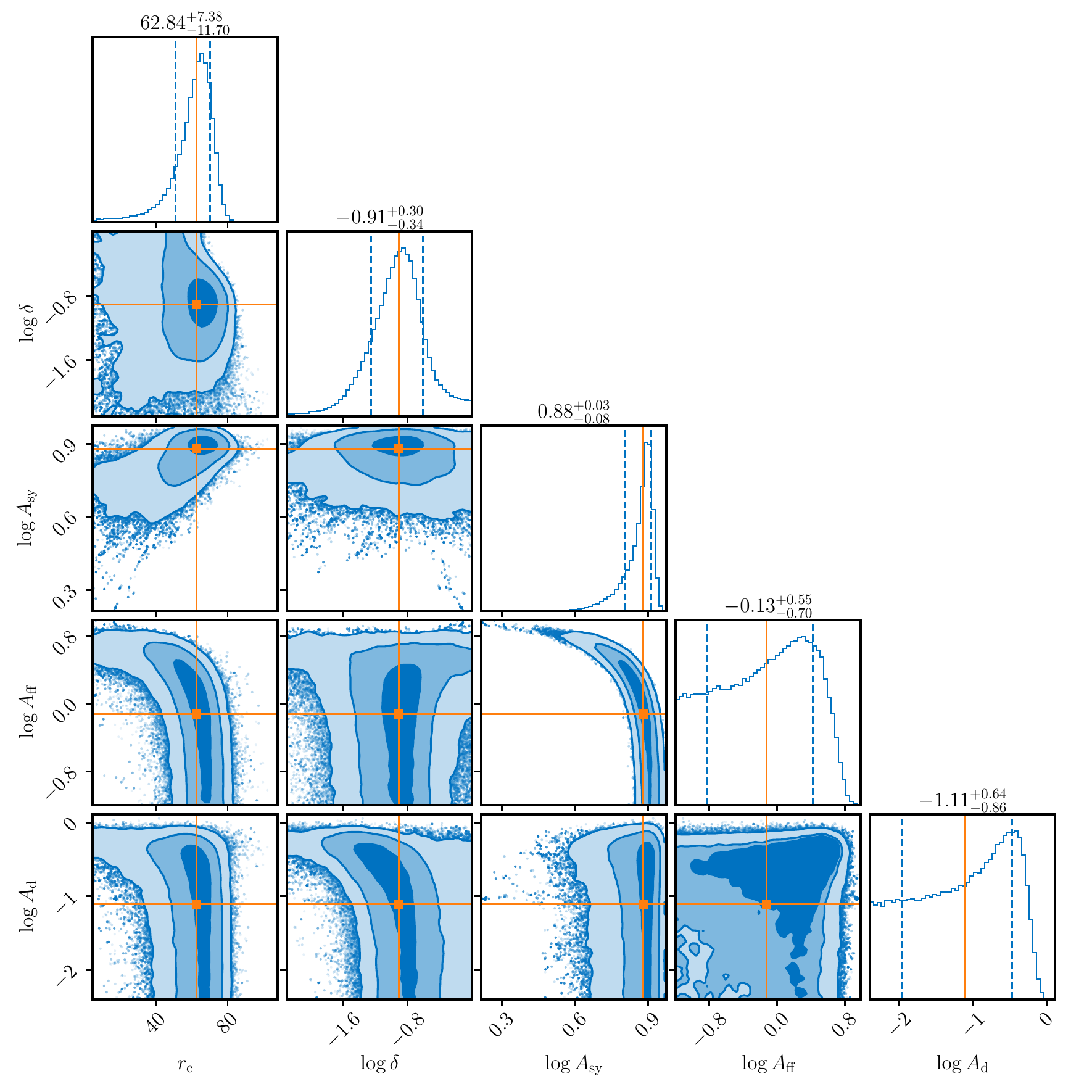} 
    \caption{Cornerplot for the posteriors of the SED fit for NGC~1068.}
    \label{fig:ngc1068_cornerplot}
\end{figure*}

\begin{figure*}
    \centering
    \includegraphics[width=\linewidth]{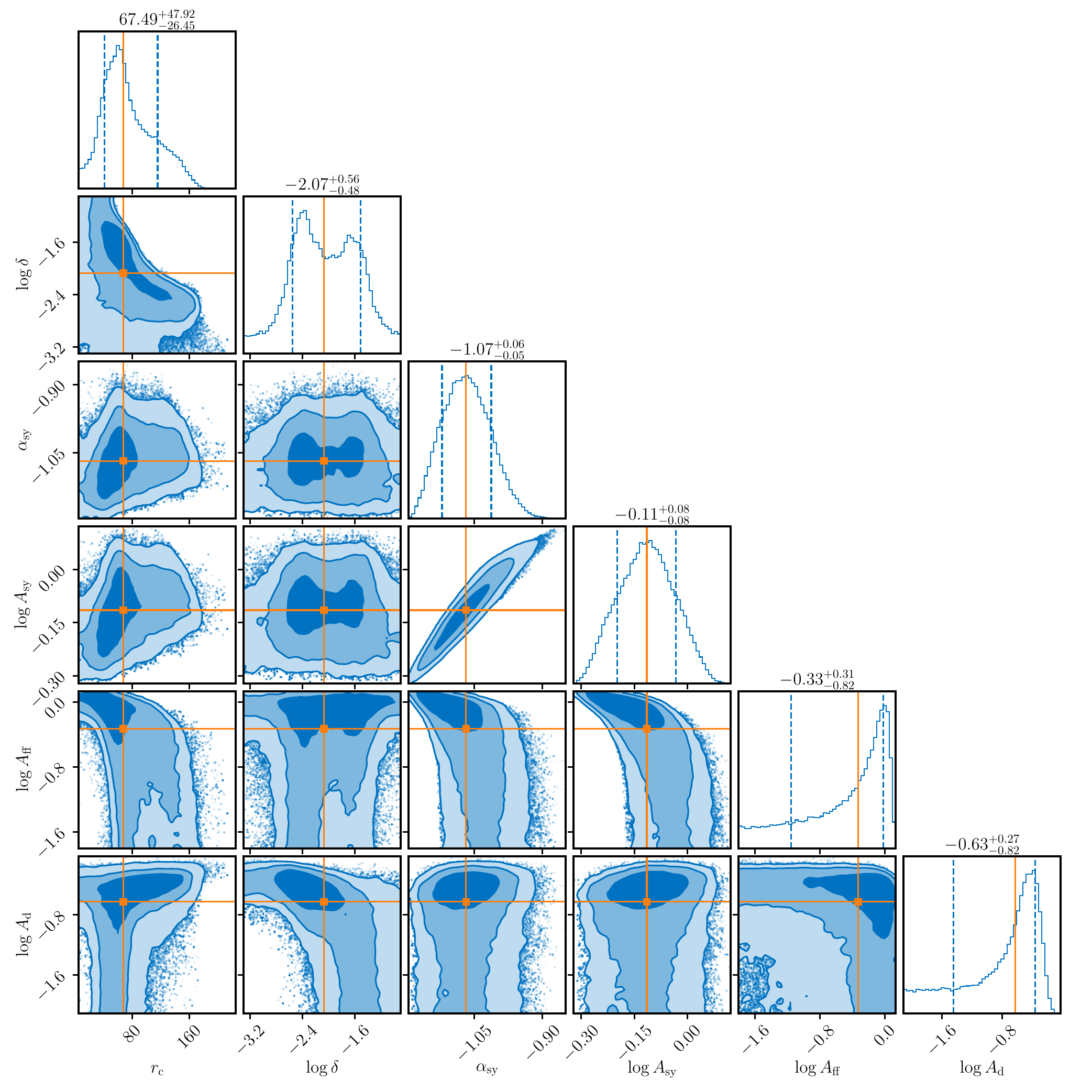} 
    \caption{Cornerplot for the posteriors of the SED fit for NGC~3227.}
    \label{fig:ngc3227_cornerplot}
\end{figure*}

We show the cornerplots for all the SED fits in Figs.~\ref{fig:grs_cornerplot}--\ref{fig:ngc3227_cornerplot}. The top panels of each plot show the 1-D distributions of the individual parameters, with the orange line marking the position of the median, and the dashed lines the 1-$\sigma$ confidence interval.

\end{document}